# A DoE-based approach for the implementation of structural surrogate models in the early stage design of box-wing aircraft

V. Cipolla [1], K. Abu Salem[2], G. Palaia[3]

*University of Pisa, Via G. Caruso 8, Pisa, Italy*

V. Binante [4], D. Zanetti[5]

*SkyBox Engineering S.r.l., Via G. Caruso 8, Pisa, Italy*

## ABSTRACT

One of the possible ways to face the challenge of reducing the environmental impact of aviation, without limiting the growth of air transport, is the introduction of more efficient, radically different aircraft architectures. Among these, the box-wing one represents a promising solution, at least in the case of its application to short-to-medium haul aircraft, which, according to the achievement of the H2020 project "PARSIFAL", would bring to a 20% reduction in terms of emitted $CO_2$ per passenger-kilometre. The present paper faces the problem of estimating the structural mass of such a disruptive configuration in the early stages of the design, underlining the limitations in this capability of the approaches available by literature and proposing a DoE-based approach to define surrogate models suitable for such purpose. A test case from the project "PARSIFAL" is used for the first conception of the approach, starting from the Finite Element Model parametrization, then followed by the construction of a database of FEM results, hence introducing the regression models and implementing them in an optimization framework. Results achieved are investigated in order to validate both the wing sizing and the optimization procedure. Finally, an additional test case resulting from the application of the box-wing layout to the regional aircraft category within the Italian research project "PROSIB", is briefly presented to further assess the capabilities of the proposed approach.

---

[1] Assistant Professor, Department of Civil and Industrial Engineering
[2] PhD Candidate, Department of Civil and Industrial Engineering
[3] PhD Student, Department of Civil and Industrial Engineering
[4] Aeronautical Engineer, PhD, Structural design and analysis
[5] Aeronautical Engineer, Aircraft design and performance analysis

# 1     Acronyms

AVL   Athena Vortex Lattice

DoE   Design of Experiment

CG  Centre of Gravity

FEM   Finite Element Model

MDO   Multi-Disciplinary Optimisation

PrP     PrandtlPlane

TLAR(s)   Top Level Aircraft Requirement(s)

VLM   Vortex-Lattice Method

# 2     Introduction

The ACARE Strategic Research and Innovation Agenda ([1]) outlines the main challenges of the future European aviation, following the line marked out by the European Commission's document Flightpath 2050 ([2]); in particular, the main objectives concern the development of a more efficient and environmentally friendly air transport. According to [1], the strategy for facing these challenges involves a great number of action areas, in which different kind of actors, such as industry, institutions, SMEs or academia, can play a major role.

Another key goal is related to the capability of fulfilling market and society demands, as it is forecasted a huge growth of the air traffic for the next decades; this aspect seems to be contrasting to the environmental challenge. To address this trade-off, the main actions are currently related to develop the next generation of air vehicles through evolutionary steps, to improve air operations, traffic management, and airport environment, to provide the necessary quantity of affordable alternative energy, and to assess aviation's climate impact.

Given this general picture, even research projects aiming at developing low TRL technologies can provide an important contribution to face the environmental challenge. This is the case of the projects in which the research here presented has been developed: a EU-funded project called "PARSIFAL" ("Prandtlplane ARchitecture for the Sustainable Improvement of Future AirpLanes"), carried out between 2017 and 2020 with the goal of studying the application of the box-wing architecture to the short-to-medium haul aircraft, and the Italian research project "PROSIB" (2018-2021), dedicated to the study of regional aircraft with hybrid-electric propulsion systems, also for the case of box-wing architectures. The box-wing configuration studied in these projects, artistically represented in Fig. 1, has been called "PrandtlPlane" (or PrP), since its development comes from the studies carried out by Ludwig Prandtl in the 1920s; Prandtl in [3] indicated the box-wing architecture as the "best wing system", i.e. the lifting system capable to minimize the induced drag for given lift and wingspan.

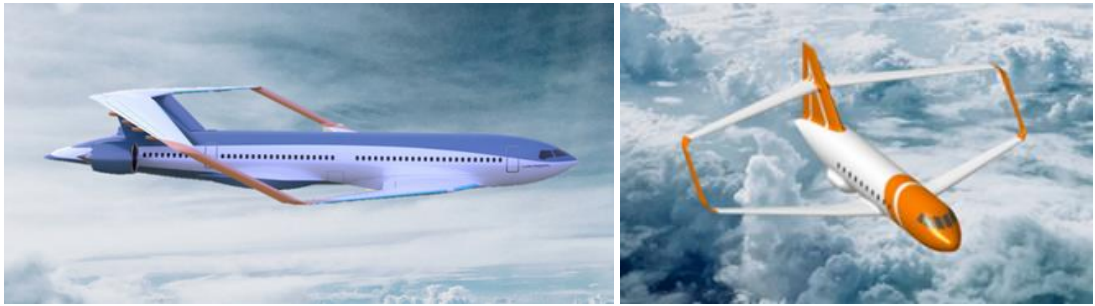

**Fig. 1. Artistic view of the PrandtlPlanes studied in the projects "PARSIFAL" (left) and "PROSIB" (right)**

The PARSIFAL project has been concluded showing that the box-wing architecture can be exploited to increase the payload capability and to reduce the fuel consumption per passenger-kilometre at the same time, compared to tube-and-wing configurations. According to PARSIFAL results, the box-wing allows to avoid the increase of induced drag that occurs when in a tube-and-wing configuration the fuselage is enlarged without increasing the wingspan. Since box-wing has a higher span efficiency, it is possible to increase the cabin width, hence the number of seats, without changing the wingspan. As summarized in [4], the results of the comparison between a PrP, designed to be compliant with the ICAO Aerodrome Reference Code "C"[6], and a conventional aircraft of the same category, represented by the common reference model called CeRAS-CSR01 ([5]), are really promising:

- up to 20% of fuel consumption and CO2 per passenger-kilometre reduction, with significant impact on both environment and market opportunities (refer to [6] and [7] for more details);

---

[6] wingspan between 24 and 36 m; outer main gear wheel span between 6 and 9 m.

- up to 18% and 23% reduction of Global Warming Potential and Global Temperature change Potential, respectively (refer to [7] for more details);

These results have been achieved after a 3 years long research in which aircraft design has played a major role. More in details, the PrP design has been performed through different phases, which can be numbered as follows:

1. a conceptual phase, described in [8] and [9], where the Top Level Aircraft Requirements (TLARs) have been translated into cabin design;

2. a preliminary phase, based on the approach proposed in [10] and implemented in a tool called *AEROSTATE*, in which a design procedure driven by a VLM-based aerodynamic optimization aims at maximizing the cruise lift-to-drag ratio under a set of flight mechanic constraints;

3. definition of a baseline PrP configuration (Fig. 2, left), as result of the preliminary design phase and starting point for following activities;

4. discipline-specific high-fidelity optimizations, whose results have been used to perform several design loops;

5. definition of an updated PrP configuration (Fig. 2, right), to be used as input for the aforementioned performance analysis.

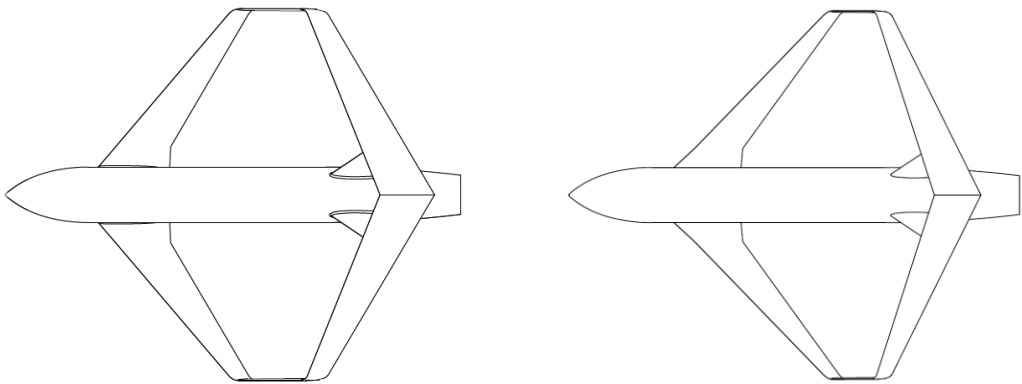

**Fig. 2. PARSIFAL project: baseline PrP configuration (left) and update after collaborative design and optimizations (right)**

Aerodynamic analyses have played a major role from the very beginning (phase "1"), in which the preliminary definition of the design space containing the feasible PrP configurations has been carried out by means of VLM analyses. On the contrary, structural analyses have been involved in the design process only after a refined design space has been available (end of phase "2") and only high-fidelity tools have been adopted from that point after.

The present paper aims to propose an approach to create and adopt surrogate structural models with the following main characteristics:

- low computational costs in order to be coupled with aerodynamic tools such as panel methods;

- derived from high-fidelity physics-based models, such as FEM, specifically created to study box-wing structures;

- capable to provide detailed information concerning mass and CG of structural components.

The project PROSIB, still ongoing while this paper is presented, is here considered as a test case for the implementation of such design approach.

## 3  State of the art of box-wing structural mass estimation

The capability of estimating the aircraft wing mass in the initial stages of the design is a well-known challenge of aircraft design. As described in [11], for conventional cantilever wing structures many approaches can be adopted, ranging from statistical methods based on previously constructed aircraft of the same type to FEM-based methods, which are usually not implementable in early design stages because of the large amount of required input data. More details on the appropriate level of detail needed to describe the structural geometry of aircraft and to achieve an acceptable level of accuracy in mass estimation is provided in [12]. According to [13], aircraft wing mass methods suitable for the conceptual design phase can be classified as in Table 1, where the main characteristics are summarized.

**Table 1. Classification of wing mass estimation methods suitable for the conceptual design phase ([13])**

| Method classification | Available Input | Achievable Output | Models implemented |
|---|---|---|---|
| Class I | TLARs | Maximum Take-off Weight (MTOW), Operational Empty Weight (OEW), Payload (PLW) and Fuel Weight (FW) | Statistical data, Basic performance equations |
| Class II | Aircraft dimensions, MTOW, FW and OEW | Mass of main aircraft components (fuselage, landing gear, wing and systems) | Semi-empirical relations based on statistical data |
| Class II+1/2 | Simplified aircraft geometry model, loads distributions and load cases definition | Detail design of primary structures, mass estimation of secondary structures | Elementary physics-based analysis for primary structures, semi-empirical and statistical methods for secondary structures |
| Class III | Detail aircraft geometry model, loads distributions and load cases definition | Detail design of structural components | Physics-based analysis, typically with FEM and CFD tools |

As said, Class III methods are seldom adopted in early design stages, since most of the time the level of details of CAD/CAE input data required to run FEM and CFD tools is too high. An example of a possible CAD-free approach is given in [14].

Moving to unconventional architecture such as the box-wing, it is well clear that the adoption of Class I and II methods are not suitable. Therefore, the main challenge faced in this work, as well as by many other authors, is to define Class II+1/2 methods specifically for box-wing structures, whose accuracy can be assessed by means of higher-fidelity methods. Concerning previous works, significant contributions can be found in the following:

- in [15], where the wing mass estimation formula proposed by Howe for cantilever wings ([16]) is adjusted for the box-wing case, introducing a corrective factor derived from a regression analysis of results obtained applying Howe's approach ([17]) to a beam model representing the wings' torsion box;

- in [18], where the authors underline the limitations of empirical models conceived to estimate the mass of cantilever wings, comparing the predictions of **Errore. L'origine riferimento non è stata trovata.** with results achieved considering the load distribution of a box-wing system;

- in [20], where a multidisciplinary optimization procedure is implemented coupling to the vortex-lattice aerodynamic solver AVL ([21]) with a FEM in which the box-wing structures are modelled using shell elements ([22]);

- in [23], which is based on a multidisciplinary optimization framework implementing the equivalent beam finite element model described in [24] to evaluate wing mass.

In addition, in [25] and [26] a multi-fidelity approach for the wing-box optimization is applied to a PrP configuration resulting from a previous aerodynamic design phase, therefore well after the early design stage.

It is also worth underlining that [15], [18] and [23] focus on the application of the proposed method to short-to-medium haul aircraft, whereas [25] and [26] address the long haul category and [20] the regional one.

To better understand how the proposed approach here presented can be integrated in the preliminary design workflows adopted for the PrP, it is worth to summarize the most relevant research which defines the state of the art for the preliminary design of the PrP:

- in [27], the preliminary design of a light sport PrP (project "IDINTOS") has been performed using the aforementioned tool *AEROSTATE*, which solves the optimization problem in Eq. (1) by calculating L/D using the VLM code AVL and considering a set of constraints - indicated as $g(x) \geq 0$ and $h(x)=0$ in Eq. (1) - on trim, stability and stall. In this case, the box-wing mass has been estimated by assuming constant surface density values, derived from statistics on existing light sport airplanes;

$$\begin{cases} min\left(-\dfrac{L(\boldsymbol{x})}{D(\boldsymbol{x})}\right)_{cruise} \\ g(\boldsymbol{x}) \geq 0 \\ h(\boldsymbol{x}) = 0 \\ \boldsymbol{x}_{min} \leq \boldsymbol{x} \leq \boldsymbol{x}_{max} \end{cases} \qquad (1)$$

- as described in [8] and [9], *AEROSTATE* is adopted in a similar way in the project PARSIFAL. During the preliminary phase of the project, the structural mass estimation has been performed assuming conservative values on MTOW mass fraction for wings and fuselage (then verified by means of FEM analyses as in [28]);

- conceptual design methods for Hybrid-Electric Propulsion (HEP) aircraft are presented in [29], focusing on the case of the PrP regional aircraft studied in the project "PROSIB". In this case, *AEROSTATE* is involved in the design workflow shown in Fig. 3, where the box-wing mass is estimated by the tool *THEA-CODE* by means of literature models valid for cantilever wings ([30]), in order to have conservative results.

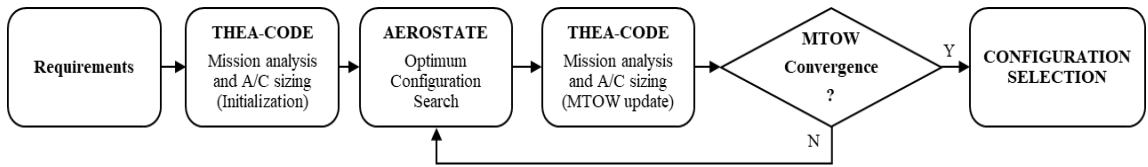

**Fig. 3. Design workflow adopted in the project "PROSIB"**

As shown, the design approaches adopted during the preliminary phases of the above cited research on the PrP has been focused mostly on aerodynamics and flight mechanics, involving low to medium fidelity physics-based tools, whereas the structural mass estimation has usually been introduced through empirical or handbook methods, in general not suitable for overconstrained wing structures as the box-wing architecture one.

It is worth underlining that the methods cited above give different importance to the several design criteria relevant for wings' structural sizing. For example, the approach proposed in **Errore. L'origine riferimento non è stata trovata.** is built on static strength criteria and other effects, such as buckling, and are not taken into account. According to [31], this simplification may be acceptable for wide-body aircraft, with wing loading above 600 kg/m², whereas it leads to significant underestimations for narrow body and regional jet aircraft.

As detailed in the following sections, although the parametrization adopted for wing-box structures would allow to model stiffened panels designed according to both strength and buckling criteria, the method here presented follows an implementation strategy driven by strength criteria only.

## 4 Approach for the development of a structural model for box-wing aircraft preliminary design

### 4.1 Parametric description of the box-wing structures

Fig. 4 shows a typical FEM model of the box-wing structure of a PrP configuration; the mesh is made of shell and beam elements and includes ribs, skins, spar webs, stringers, and spar caps. Shell elements are used for modelling skins, ribs and spar webs, whereas stringers and spar caps are modelled as beam elements. As shown, the FE mesh involves only the wing-box structure of the wings, whereas the fixed and movable parts of both leading and trailing edge are modelled as point masses attached to the spar webs, by means of surface-based constraint relationships.

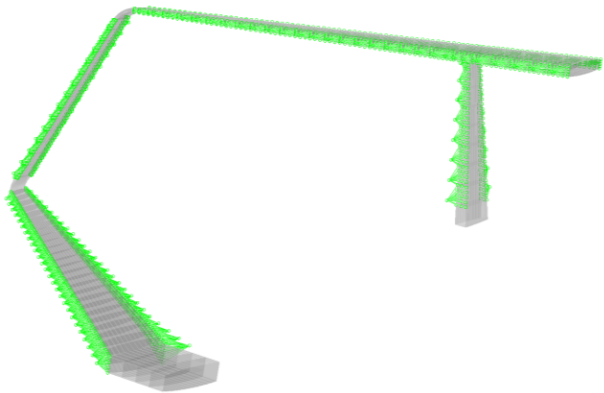

**Fig. 4. A typical FEM model of the box-wing of a PrP configuration; wing-box structure depicted in grey, whereas point masses of the LE and TE edge sections are linked to spar webs**

For each wing of the box-wing system, the overall structural parameters are shown in Table 2 and Fig. 5; most of them are related to a given wing section, whereas their spanwise change is expressed by means of the stringer tip-to-root scale factor $\alpha_{st}$, which is introduced assuming that the stringer cross-section does not change its shape along the wingspan, and the tip-to-root thickness ratio of the wing-box panels, which are defined for top and bottom skins ($\tau_{sk}$) and lateral webs ($\tau_{web}$).

**Table 2. Full set of structural parameters for the box-wing system**

| Wing-box components | Structural parameters | Description |
|---|---|---|
| **Spar caps** | $w_{\mathrm{sp}}^{\mathrm{i}}$ | Spar cap width at root section |
| | $h_{\mathrm{sp}}^{\mathrm{i}}$ | Spar cap height at root section |
| | $t_{\mathrm{sp}}^{\mathrm{i}}$ | Spar cap thickness at root section |

| **Stringers** | $w_{st}^{ij}$ | Stringer width at root section |
|---|---|---|
| | $t_{st}^{i}$ | Stringer thickness at root section |
| | $p_{st}^{i}$ | Stringer pitch (constant along wingspan) |
| | $h_{st}^{ij}$ | Stringer height at root section |
| | $\alpha_{st}^{i}$ | Stringer tip-to-root scale factor |
| **Spar webs** | $t_{web}^{i}$ | Spar web thickness at root section |
| | $\tau_{web}$ | Web thickness tip-to-root thickness ratio |
| **Skins** | $t_{sk}^{ij}$ | Skin thickness at root section |
| | $\tau_{sk}^{i}$ | Skin tip-to-root thickness ratio |
| **Ribs** | $t_{rib}^{i}$ | Rib thickness (constant along wingspan) |
| | $p_{rib}^{i}$ | Rib pitch (constant along wingspan) |

**Notes: Superscript *i* indicates the wing (F=front wing, R= rear wing, V= vertical tip-wing, T=vertical tail plane). The superscript *j* indicates the top (T) or the bottom (B) wing panel.**

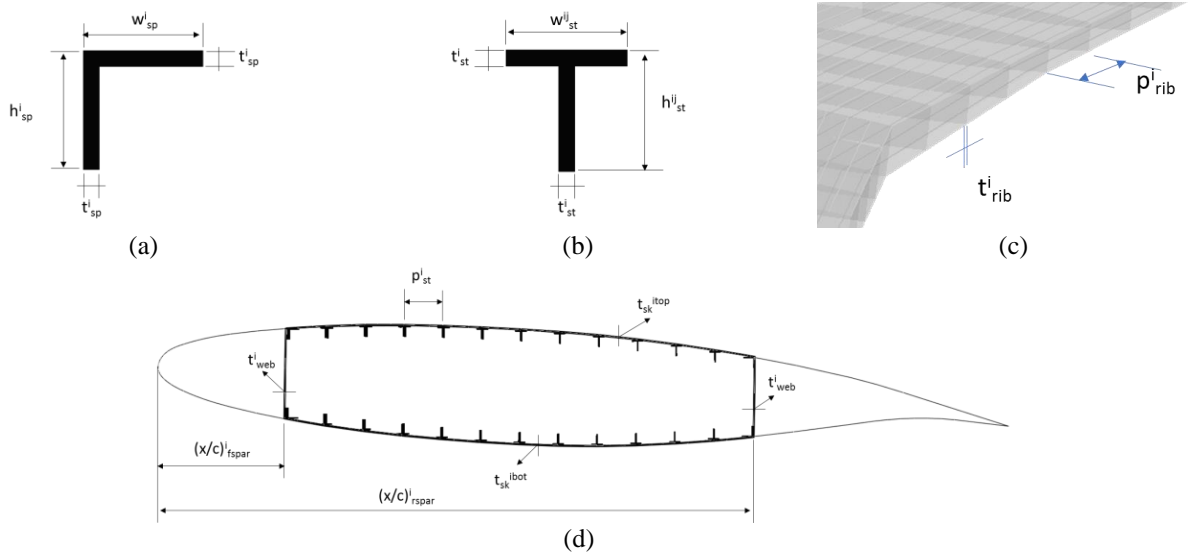

(a)　　　　　(b)　　　　　(c)

(d)

**Fig. 5. Structural parameters of a wing of the box-wing structure: (a) spar caps; (b) stringers; (c) ribs; (d) wing-box cross-section**

Therefore, the following relationships describe any stringers' dimension ($f_{st}^{ij}$) and the wing-box panels thickness along the wingspan, for a straight tapered wing:

$$f_{st}^{ij}(y) = \left( \frac{(1-\alpha_{st}^i) \cdot y + \alpha_{st}^i \cdot y_{root}^i - b/2}{y_{root}^i - b/2} \right) \cdot f_{st}^{ij} \big|_{root} \tag{2}$$

$$t_{sk}^{ij}(y) = \left( \frac{(1-\tau_{sk}^i) \cdot y + \tau_{sk}^i \cdot y_{root}^i - b/2}{y_{root}^i - b/2} \right) \cdot t_{sk}^{ij} \big|_{root} \tag{3}$$

$$t_{web}^i(y) = \left( \frac{(1 - \tau_{web}^i) \cdot y + \tau_{web}^i \cdot y_{root}^i - b/2}{y_{root}^i - b/2} \right) \cdot t_{web}^i \big|_{root} \tag{4}$$

where $b$ is the wingspan and all the other symbols have the same meaning than in Table 2. It is worth to note that $\alpha_{st}$ and $\tau_{sk}$ can refer both to top and bottom panels, whereas thickness root values can be different.

With reference to Table 2, each wing-box structure is characterized by 17 parameters, 11 of which are related to the wing root section. It is worth noting that the number of parameters could be higher if we assume different stringer thicknesses between the upper and lower skin, as well as different spar web thicknesses between the front and the rear spar and different spar cap's parameters between the upper and lower skin and between the front and the rear spar. Since the DoE size, i.e. the number of parameters' combinations to be evaluated, is $2^n$ in a two-level full factorial design based on $n$ global parameters, further simplifications have been introduced in order to reduce the number of design parameters.

Focusing on structural sizing of front and rear wings, the structural parameters of vertical wings and vertical tail plane are fixed to values obtained in [32] and more detailed in [33]; although the number of structural parameters has been reduced to that of the horizontal wings this is still high for the generation of a DoE, regardless of the design to be considered (e.g. two-level full factorial, or a central composite faced design, or three-level partial factorial, and so on), requiring a very high number of FEM simulations.

To further reduce the structural design variables, it has been decided to keep some of the main wings' parameters constant; in particular, on the basis of previous structural optimizations of both baseline and updated PrP configuration of Fig. 2, we decided to keep the following parameters to their optimal values, that is:

- stringer's pitch has been fixed to about 155 mm and 170 mm for front and rear wing, respectively;

- the ratio between the stringer's height and width has been kept constant to 1 for both wings; the same ratio has been used for spar caps;

- for both wings, spar cap thickness has been chosen equal to stringer thickness, whereas an average value of stringer heights between upper and lower skins has been used for spar caps height;

- ribs' pitch has been fixed to about 600 mm, while the rib thickness has been fixed to 4 mm;

- according to [32] and [33], the skin thickness ratio for the front wing seems to have a little influence on structural design, therefore it has been fixed to the value of 0.295 as done in these works;

- similarly, the spar web thickness of the front wing and that of the rear wing have been fixed to the optimal values;

- for the sake of simplicity, the same thickness ratio value is chosen for skins and webs, hence introducing the following assumption for any wing-box of the system:

$$\tau_{web}^i = \tau_{sk}^i \qquad (5)$$

- again, from the above-mentioned works, some structural parameters of the rear wing seem to have a little influence on the optimal solution, whereas some other ones have been set to appropriate values to provide enough stiffness to the wing itself. Such parameters are the thickness of the lower skin, the stringer's thickness, the height of the stringers of the bottom panel and the stringer tip-to-root scale factors; therefore, they have been fixed to their optimal value.

On the basis of the above assumptions, the number of structural parameters is reduced to 9 as shown in Table 3.

**Table 3. Reduced set of structural parameters for the box-wing system**

| Design parameter | Description |
|:---:|:---|
| $t_{sk}^{FB}$ | Front wing / Bottom panel / Skin thickness |
| $t_{sk}^{FT}$ | Front wing / Top panel / Skin thickness |
| $t_{st}^{F}$ | Front wing / Stringers thickness |
| $h_{st}^{FT}$ | Front wing / Top panel / Stringers height |
| $h_{st}^{FB}$ | Front wing / Bottom panel / Stringers height |
| $\alpha_{st}^{F}$ | Front wing / Stringer scale factor (tip/root) |
| $h_{st}^{RT}$ | Rear wing / Top panel / Stringers height |
| $t_{sk}^{RT}$ | Rear wing / Top panel / Skin thickness |
| $\tau_{sk}^{R}$ | Rear wing / Skin and web thickness ratio (tip/root) |

### 4.2    FEM database construction

Regardless of the type of DoE one wants to consider, sensitivity analysis involves hundreds or thousands of FEM analyses, each of which associated with a given set of values of the structural parameters of Table 3. For such analyses, an in-house tool called WAGNER ([28]) has been used; it is a python tool for automatic generation of 3D structure models and finite element mesh of PrandtlPlane configurations, and, thus, particularly suitable for sensitivity analyses. By using standard libraries of the Abaqus commercial FEM software [34], WAGNER performs global structural analyses.

The sensitivity analysis is carried out by keeping constant the geometry of a box-wing configuration taken as a reference, while varying the structural variables. The reference box-wing configuration is here called "PrP-300", which has been obtained as a refinement of the final design resulting from the project PARSIFAL ([35]). Table 4 shows the Top Level Aircraft Requirements (TLARs) and its overall dimensions.

**Table 4. TLARs and dimensions of the reference box-wing aircraft "PrP-300"**

| | |
|---|---|
| **Max n° passengers** | 308 |
| **Seat abreast** | 2-4-2 |
| **Design Range** | 2700 nm |
| **Cruise altitude** | FL 360 |
| **Cruise Mach** | 0.79 |
| **Ref. Wing Area** | 266.7 m$^2$ |
| **Wingspan** | 36 m |
| **Fuselage length** | 44.2 m |
| **MTOW** | 128688 kg |

Fig. 6 shows a general overview of the FE model of half a structure of the PrP-300 configuration, with a particular focus on the parametrization of the wing-box structure of the main wings; all design variables at root section of each wing are shown.

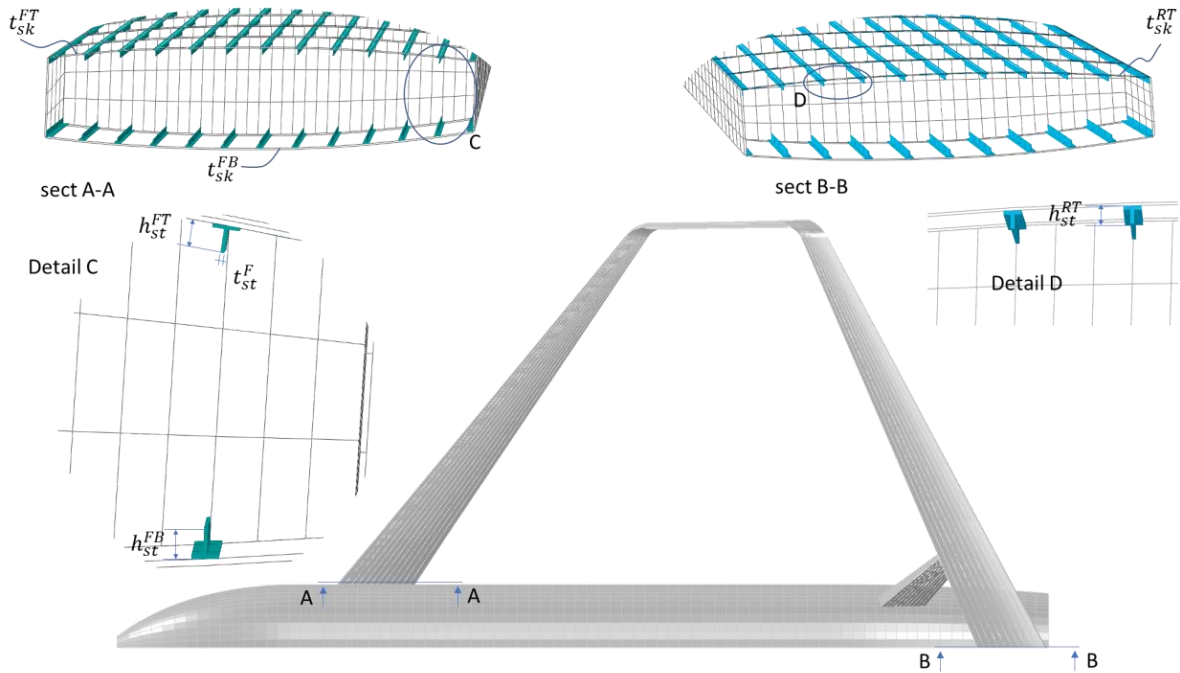

**Fig. 6. Overview of the FE model of half a structure of the PrP-300 configuration, with some details about wing-box structure parametrization**

A detailed FE model of the PrP-300 configuration is shown in Fig. 7. The finite element mesh involves the wing-box structure and the fuselage body delimited by nose and tail bulkhead; as already mentioned, it is mainly made of shell and beam elements and includes all primary structural components. Secondary structures, such as leading edge (LE) and trailing edge (TE) movable surfaces, engines and landing gears, are modelled as point masses linked to primary structures; in particular, as shown in Fig. 7, concentrated masses of the fixed and movable structures of both LE and TE sections are connected to the spar's webs of the wing-box structure. The estimation of such masses has been performed according to [36]. Engines as well as landing gears are linked to main frames and stringers of the fuselage body. All the connections among primary and secondary structures and those related to the attachment between two primary structures (such as fuselage-front wing connection and so on) are modelled as surface-based constraints which provide a more realistic constraint type (with respect to multipoint constraints or node-based constraints) and more accurate stresses deriving from the connection.

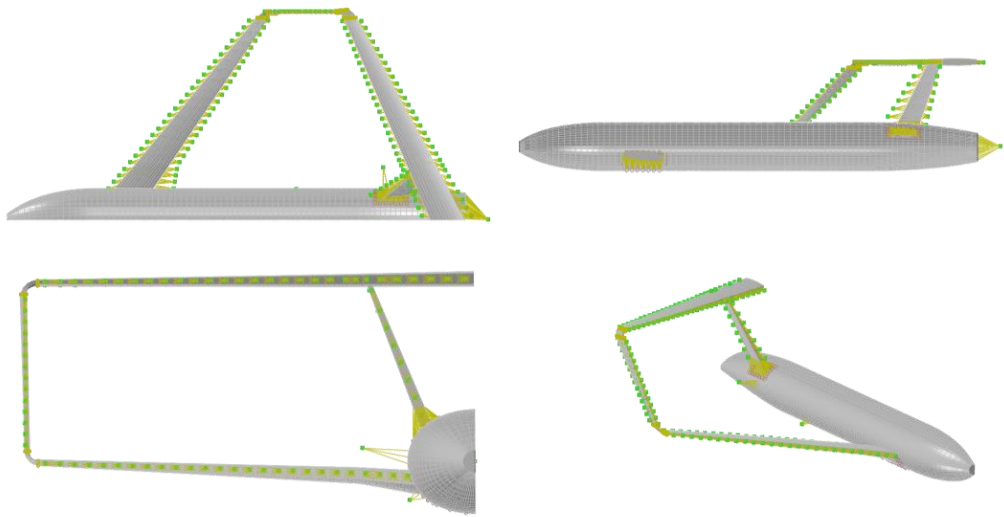

**Fig. 7. Detailed FE model of half a structure of the PrP-300 configuration**

In addition to the weight of primary and secondary A/C structures, the FE model accounts for miscellaneous masses related to systems, equipment and other non-structural masses, evaluated according to [37]; these masses have been added to the FE model as non-structural masses spread across the fuselage structure. Payload has been modelled as non-structural mass applied to the deck floor beams; fuel mass has also been modelled as non-structural masses added to the wing-box structure where fuel tanks are located. The FE model so built includes the maximum take-off weight of half a structure.

For structural sizing of a PrP configuration several load cases as combinations of pressurization loads, aerodynamic forces and inertial loads can be considered; the inertial loads can be used to account for engine thrust and forces introduced by landing gears at landing. Fuselage has been designed for carrying out the pressurization load and inertial loads deriving from $n_z$=+2.5. Keeping in mind that the main objective of the sensitivity analysis is to create a DoE for sizing the horizontal wings of a box-wing structure, the fuselage as well as vertical tip-wing and vertical tail plane have been preliminarily sized and their structure kept unchanged in the sensitivity analysis. The sizing criterion adopted to preliminary size these structures is similar to that used for the wing-box structure of the horizontal wings, discussed later.

Concerning the wing-box structure, the sizing loading condition considered in this work is associated to a symmetric load case related to a combination of aerodynamic forces and gravity loads; aerodynamic forces are modelled as nodal forces applied on the pressure centre of each wing-bay, starting from the spanwise lift distribution in take-off condition evaluated by the AVL code (see Fig. 8). A gravity load related to $n_z$=+2.5 has been superimposed to the whole FE model. Fig. 8 also shows the boundary conditions of longitudinal symmetry.

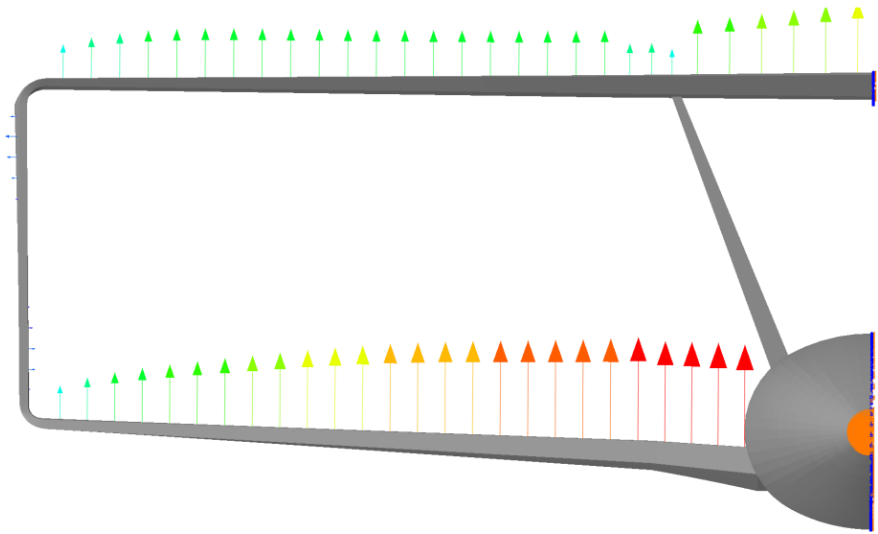

**Fig. 8. Aerodynamic forces obtained from AVL code and applied to the wing-box structure of the reference PrP configuration**

For the above load cases the sizing criterion is based on the requirements that the maximum calculated stress is not greater than the admissible one and the deformations of the box-wing structure must be in the range of geometric linearity. The admissible stress is evaluated as the yield strength divided by an appropriate safety factor. Concerning the materials used for the current FE analyses, aluminium alloys have been considered; in particular, Al-2024 for skins, spar caps and stringers, whereas Al-7075 for ribs and spars web. Regarding the geometric linearity condition, it has been assumed that the maximum wing-tip displacement should not exceed 10% of the wing semi-span.

### 4.3   The DoE approach

The DoE approach is based on sensitivity analyses which aim to build regression models for the prediction of the wing structural mass, the estimation of the stress in each main wing and the evaluation of the highest wing-tip displacement between front and rear wings. To this end, the structural design variables have been varied within ranges shown in Table 5; the choice of such ranges of variation has been made on the basis of previous results reported in [32].

**Table 5. Design space and ranges of variation for DoE**

|  | Design variables | Lower bound [mm] | Upper bound [mm] | Step size [mm] | Codified variables |
|---|---|---|---|---|---|
| **Front Wing** | $t_{sk}^{FT}$ | 6 | 14 | 4 | x1 |
|  | $t_{sk}^{FB}$ | 6 | 14 | 4 | x2 |
|  | $t_{st}^{F}$ | 1.5 | 7 | 2.75 | x4 |
|  | $h_{st}^{FT}$ | 40 | 70 | 15 | x5 |
|  | $h_{st}^{FB}$ | 40 | 70 | 15 | x6 |
|  | $\alpha_{st}^{F}$ | 0.7 | 1 | 0.15 | x8 |
| **Rear Wing** | $t_{sk}^{RT}$ | 6 | 14 | 4 | x9 |
|  | $h_{st}^{RT}$ | 40 | 70 | 15 | x13 |
|  | $\tau_{sk}^{R}$ | 0.25 | 0.49 | 0.12 | x15 |

Table 5 also shows the codified variables $x_l$, usually used in DoE analyses ([38]) and associated with the design variables as follows:

$$x_l = \frac{v_l - \left(\frac{lb + ub}{2}\right)}{\delta} \qquad (6)$$

where $v_l$ represents the $l^{th}$ design variable, *lb* and *ub* are the lower and upper bound of its range of variation and $\delta$ is the step size by means of which the $l^{th}$ codified variable ranges from -1 to 1. In this work the codified variables can assume values ±1, thus defining two-level sensitivity design.

Concerning the choice of the type of DoE, a 2-level full factorial design has been initially considered, but the regression models derived from it are linear and, thus, are not adequate in the presence of quadratic behaviour of the response surface ([38]). This is the case for the response surfaces of stress and wing-tip displacement prediction, whereas the mass response surface does not depend on pure quadratic terms. Therefore, it has been decided to supplement the full factorial design with another design containing pure quadratic terms of the codified variables. This allows us to explore structural solutions even around the centroid of the design space where we found evidence of a curvature of the response surfaces.

The quadratic design chosen to complement the 2-level full factorial design is the Box-Wilson central composite faced (CCF) design ([39]); it is worth noting that such a design is less accurate than the central composite circumscribed (CCC) one, but the latter is not compatible with the ranges of variation defined in Table 5 and originally chosen for the full factorial design.

The CCF design requires 538 evaluations, whose $2^9=512$ are those of the 2-level full factorial design, 8 evaluations are in the centroid of the design space and 18 are on the faces of the design space. Considering that the evaluations are represented by numerical simulations, the eight evaluations at the centroid of the design space reduce to 1; therefore, sensitivity analysis requires 531 FEM simulations.

Fig. 9 shows some characteristic DoE plots related to the stress response in the front wing; Fig. 9(a) – (b) show the effect of each factor, whereas Fig. 9(c) – (d) account for the effects of the most significant 2-factor interactions.

The effects of multi-factor interactions have been also analysed, although non shown in previous pictures; the most significant interactions are between 2 or 3 factors, whereas those among 4 or more factors have little effect on the stress response. This is represented in detail in Fig. 9-d, which illustrates the sensitivity to factors or their combinations, ordered from the most to the less significant. The sensitivity to the $l^{th}$ codified variable is defined as in Eq. (7), where $\sigma_{VM}$ is the maximum Von Mises stress values detected in the front wing.

$$\Sigma_l = \frac{\left| \sum \sigma_{VM}(x_l = -1) - \sum \sigma_{VM}(x_l = 1) \right|}{2^{n-1}} \tag{7}$$

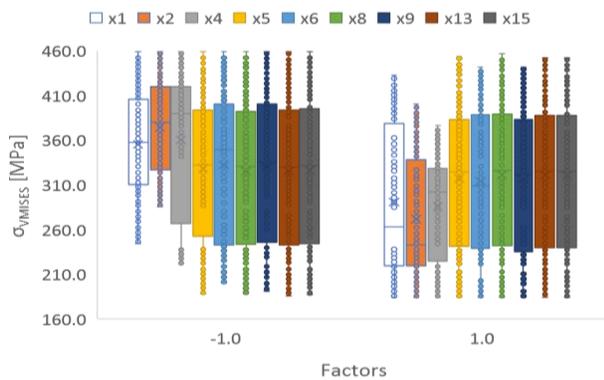

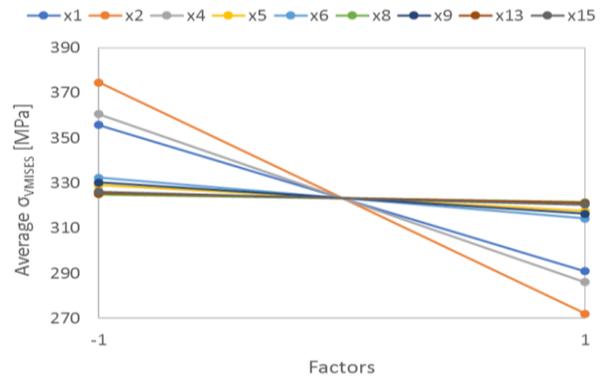

(a)

(b)

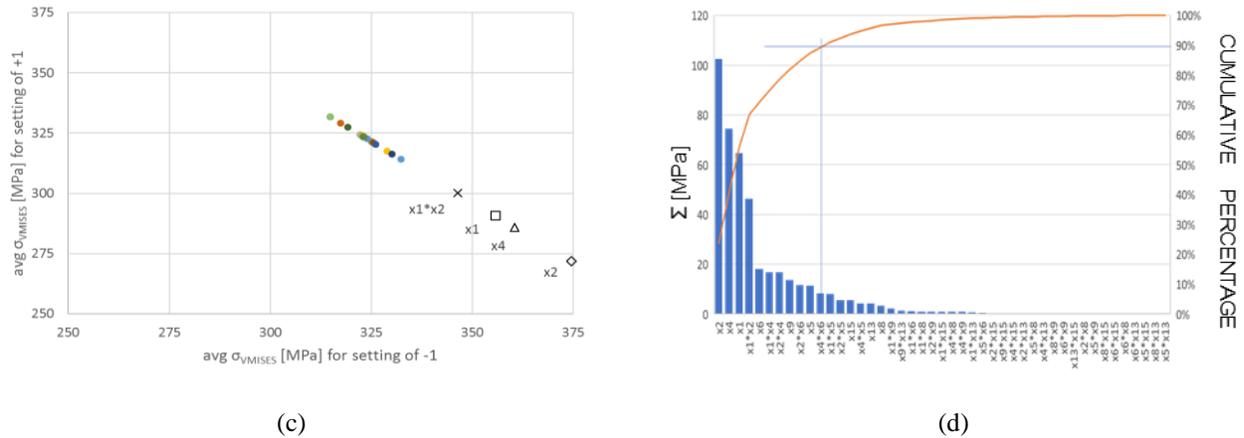

(c)                                          (d)

**Fig. 9. DoE plots for the stress in the front wing: (a) scatter plot; (b) mean plot; (c) Youden plot; (d) Pareto chart**

The analysis of DoE data leads to conclusion that only a few design variables significantly influence the stress response in the front wing; in particular with reference to Table 6, the most important variables are the skins and stringers thickness. Moreover, the thickness of the bottom skin has the highest effect among the aforementioned variables; this is clear, considering that the bending stress in the wing implies the maximum tensile stress in the bottom skin. The other parameters have little effect.

**Table 6. Effect of the design variables on the stress response in the front wing**

| Parameter | $x_2$ | $x_4$ | $x_1$ | $x_6$ | $x_9$ | $x_5$ | $x_{15}$ | $x_{13}$ | $x_8$ |
|---|---|---|---|---|---|---|---|---|---|
| $\Sigma$ [MPa] | 102.6 | 74.6 | 64.8 | 18.2 | 13.9 | 11.4 | 5.7 | 4.3 | 3.5 |
| $\Sigma/\Sigma_{max}$ [%] | 100% | 73% | 63% | 18% | 14% | 11% | 6% | 4% | 3% |
| Sensitivity level | High | Intermediate | Intermediate | Low | Low | Low | Low | Low | Low |

Fig. 10 shows the effects of the most significant factors, arranged in decreasing order; it is similar to Fig. 9(d) with the exception of including the effects of 3-factor interactions and limiting to the most important effects, according to a 90% numerical significance criterion for which factors having an effect such that $\Sigma/\Sigma_{max} < 0.1$ are considered unimportant. It is worth noting that the last three factors in Fig. 10 can be considered unimportant according to the above criterion; however, it has been decided to consider them, in order to include the effect of all structural variables on the stress response.

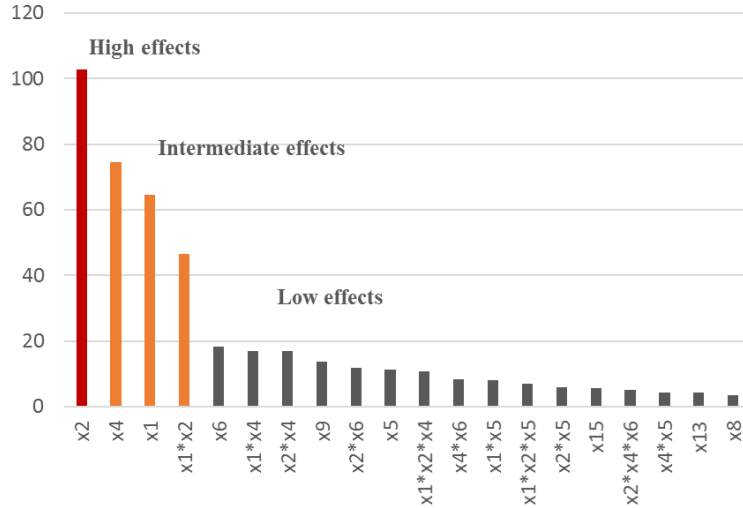

**Fig. 10. Ordered effects of the most significant factors (and combinations) for the stress response in the front wing**

From the analysis of DoE data, a regression model for the stress response in the front wing has been built by means of the software R ([40]), whose model coefficients are shown in Table 7; as shown the model involves 23 terms, including the most significant pure quadratic terms as well. The latter are related to the codified variables $x_2$ and $x_4$, that is the bottom skin thickness and the stringers thickness.

**Table 7. Coefficients of the regression model for the stress response in the front wing**

| Model coefficients | | | | | |
|---|---|---|---|---|---|
| $a_0^F$ | 269.438 | $b_{13}^F$ | -2.167 | $c_{26}^F$ | 5.84 |
| $b_1^F$ | -32.289 | $b_{15}^F$ | -2.853 | $c_{44}^F$ | 13.973 |
| $b_2^F$ | -51.358 | $c_{12}^F$ | -23.2 | $c_{45}^F$ | -2.199 |
| $b_4^F$ | -37.248 | $c_{14}^F$ | 8.455 | $c_{46}^F$ | -4.192 |
| $b_5^F$ | -5.684 | $c_{15}^F$ | 4.076 | $d_{124}^F$ | 5.426 |
| $b_6^F$ | -9.092 | $c_{22}^F$ | 39.84 | $d_{125}^F$ | 3.453 |
| $b_8^F$ | -1.733 | $c_{24}^F$ | 8.413 | $d_{246}^F$ | 2.582 |
| $b_9^F$ | -6.887 | $c_{25}^F$ | -2.883 | -- | -- |

With reference to Table 7, the regression model for the stress response in the front wing is given by:

$$\sigma_{VM}^F(\boldsymbol{x}) \cong a_0^F + \sum b_l^F \cdot x_l + \sum c_{lm}^F \cdot x_l \cdot x_m + \sum d_{lmn}^F \cdot x_l \cdot x_m \cdot x_n \qquad (8)$$

with the stress expressed in MPa. For such a model, the residual standard deviation is about 4.415, whereas the mean residual is about -2.758e-5; the goodness of the fitted model is shown in Fig. 11, showing the distribution of the residuals, evaluated as the difference between the stress calculated by FEM solver and those predicted with the regression model, and the Q-Q plot; as shown, the predicted quantiles are nearly equal to the calculated quantiles, indicating that the goodness of the model is high.

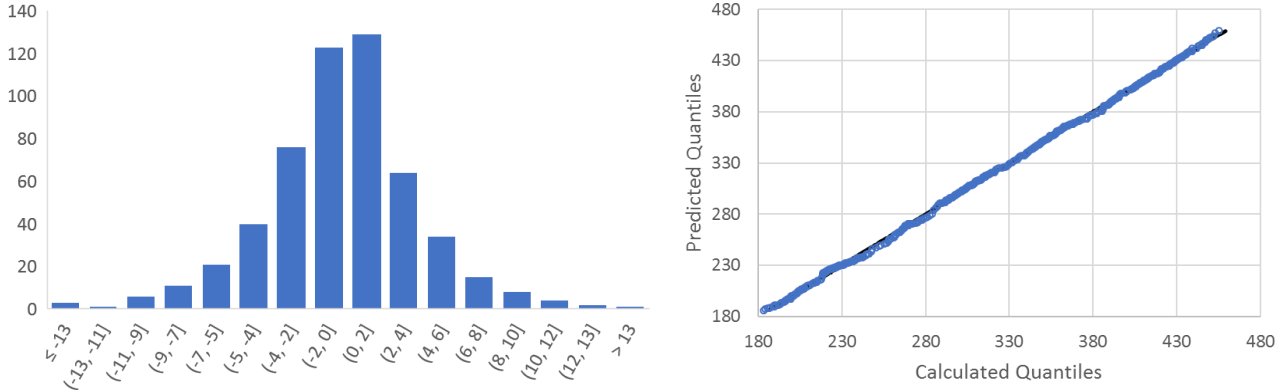

**Fig. 11. Goodness of the fitted model for the stress in the front wing: residual distribution (left) Q-Q plot (right)**

Concerning the stress response in the rear wing, the analysis of DoE data leads to conclusions similar to those drawn previously; only a few variables have a significant effect as shown in Fig. 12 and Fig. 13. In particular, the most important variables are the thickness of both skins and the stringer thickness of the front wing; the thickness of the top skin, the stringers height and the skin and web thickness ratio of the rear wing are significant as well. All these variables have comparable effects, whereas all other effects shown in Fig. 13 are low. It should be noted that, due to the over-constrained nature of a box-wing structure, the front wing variables influence the stress response in the rear wing, with an effect which is similar to that of the rear wing variables.

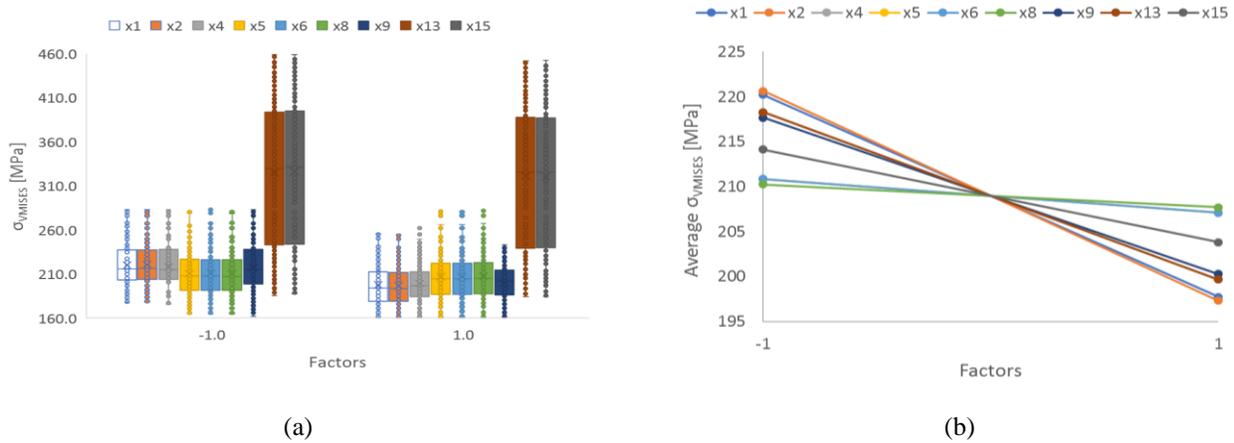

(a)                                                                 (b)

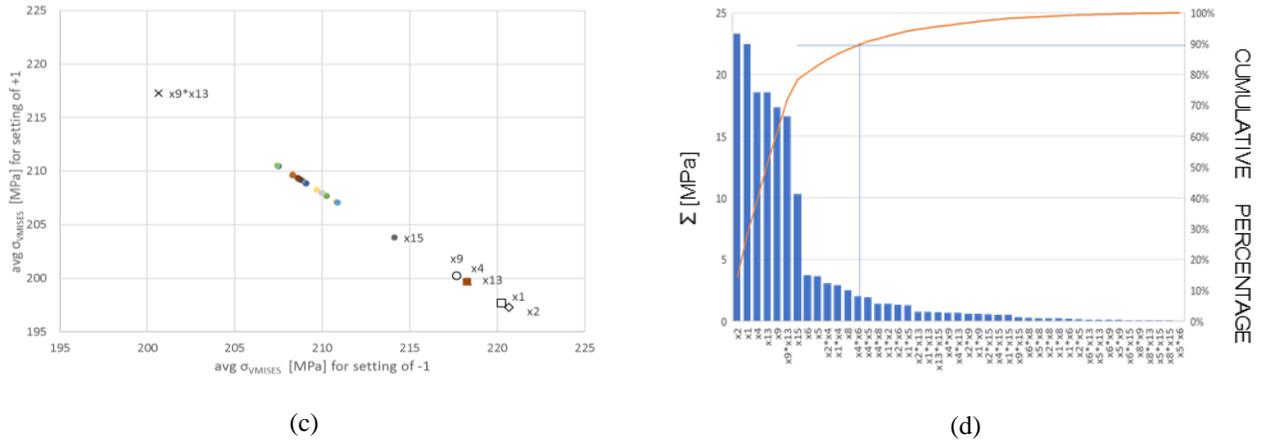

(c)                                                          (d)

**Fig. 12. DoE plots for the stress response in the rear wing: (a) scatter plot; (b) mean plot; (c) Youden plot; (d) Pareto chart**

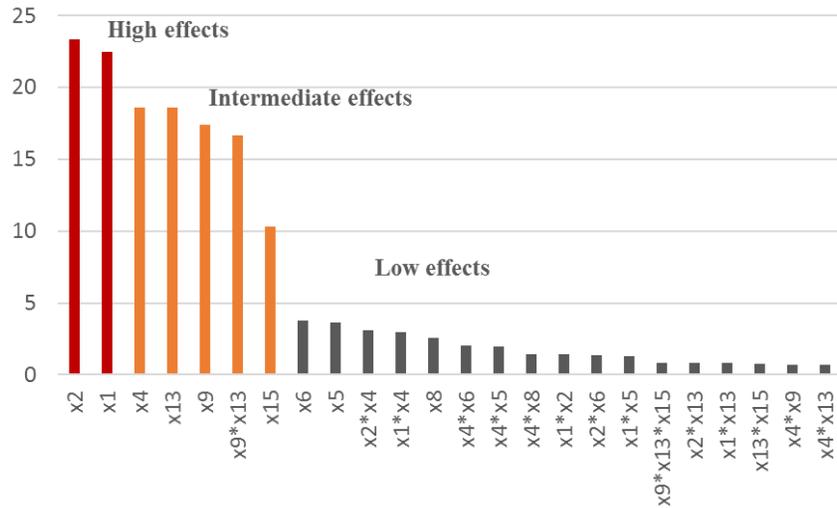

**Fig. 13. Ordered effects of the most important factors for the stress in the rear wing**

The regression model for the stress response in the rear wing is given by:

$$\sigma_{VM}^R(x) \cong a_0^R + \sum b_l^R \cdot x_l + \sum c_{lm}^R \cdot x_l \cdot x_m + \sum d_{lmn}^R \cdot x_l \cdot x_m \cdot x_n \tag{9}$$

with the model coefficients shown in Table 8. The model includes 27 terms, whose two are pure quadratic terms ($x_2^2$ and $x_9^2$), one term provides a 3-factor interaction effect and the others are related to single factors and 2-factor interactions. The stress evaluated by Eq. (9) is expressed in MPa.

**Table 8. Coefficients of the regression model for the stress response in the rear wing**

| Model coefficients | | | | | |
|---|---|---|---|---|---|
| $a_0^R$ | 193.116 | $b_{15}^R$ | -5.157 | $c_{45}^R$ | -0.99 |
| $b_1^R$ | -11.219 | $c_{12}^R$ | -0.712 | $c_{46}^R$ | -1.027 |
| $b_2^R$ | -11.667 | $c_{14}^R$ | 1.479 | $c_{48}^R$ | -0.714 |
| $b_5^R$ | -9.29 | $c_{15}^R$ | 0.663 | $c_{49}^R$ | 0.356 |
| $b_6^R$ | -1.832 | $c_{113}^R$ | 0.401 | $c_{413}^R$ | 0.351 |
| $b_8^R$ | -1.879 | $c_{22}^R$ | 5.141 | $c_{99}^R$ | 10.71 |
| $b_9^R$ | -1.267 | $c_{24}^R$ | 1.541 | $c_{913}^R$ | 8.322 |
| $b_{13}^R$ | -8.687 | $c_{26}^R$ | 0.679 | $c_{1315}^R$ | 0.369 |
| | -9.252 | $c_{213}^R$ | 0.401 | $d_{91315}^R$ | -0.429 |

For this model, the residual standard deviation is about 1.355, whereas the mean residual is about -3.567e-6; the goodness of the model is shown in Fig. 14.

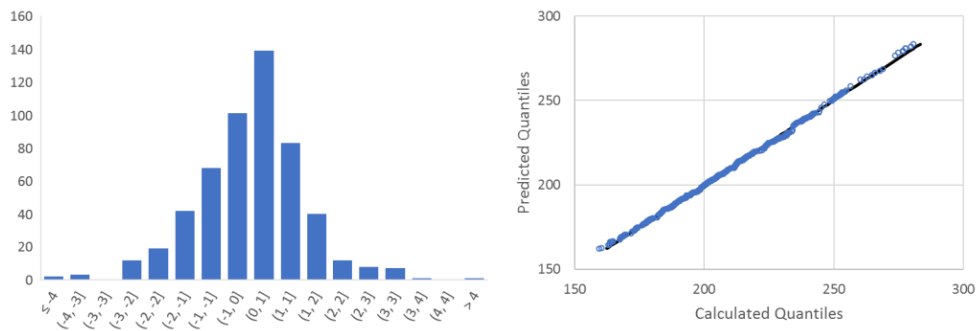

**Fig. 14. Goodness of the fitted model for the stress in the rear wing: residual distribution (left) and Q-Q plot (right)**

The regression model for the wing-tip displacement response has been built with a similar procedure used for the previous ones, starting from the analysis of DoE data; such a model is expressed as follows:

$$u_z^{max}(x) \cong A_0 + \sum B_l \cdot x_l + \sum C_{lm} \cdot x_l \cdot x_m + \sum D_{lmn} \cdot x_l \cdot x_m \cdot x_n \qquad (10)$$

where the model coefficients are shown in Table 9. The model includes 30 terms, whose three are pure quadratic terms (i.e. $x_2^2$, $x_4^2$ and $x_9^2$), two are related to 3-factor interaction effects and the others refer to single factors and 2-factor interaction effects. The wing-tip deflection given by Eq. (10) is expressed in millimetres.

**Table 9. Coefficients of the regression model for the wing-tip displacement response**

| Model coefficients | | | | | |
|---|---|---|---|---|---|
| $A_0$ | 1525.664 | $C_{12}$ | -4.283 | $C_{44}$ | 29.367 |
| $B_1$ | -141.723 | $C_{14}$ | 23.995 | $C_{45}$ | -12.108 |
| $B_2$ | -144.03 | $C_{15}$ | 9.161 | $C_{46}$ | -11.952 |
| $B_4$ | -123.96 | $C_{19}$ | 8.851 | $C_{48}$ | -8.667 |
| $B_5$ | -24.495 | $C_{115}$ | 3.687 | $C_{49}$ | 8.121 |

| $B_6$ | -23.904 | $C_{22}$ | 46.656 | $C_{415}$ | 3.516 |
|---|---|---|---|---|---|
| $B_8$ | -19.053 | $C_{24}$ | 23.303 | $C_{99}$ | 31.291 |
| $B_9$ | -90.84 | $C_{26}$ | 8.747 | $C_{913}$ | 9.483 |
| $B_{13}$ | -28.513 | $C_{29}$ | 9.265 | $D_{124}$ | 3.56 |
| $B_{15}$ | -37.948 | $C_{215}$ | 3.892 | $D_{246}$ | 3.716 |

With reference to Table 9, the most important factors for the wing-tip displacement response are the skins thickness of both wings and the stringers thickness of the front wing, to which the coefficients B1, B2, B9 and B4 are associated. Pure quadratic terms in x2, x4 and x9, related to the coefficients C22, C44 and C99 have intermediate effects, as well as the stringers height and the skin and web thickness ratio of the rear wing to which the coefficients B13 and B15 are associated. The other factors, including multiple factor interactions have low-moderate effect on the wing-tip displacement response.

The goodness of the fitted model is shown in Fig. 15; the residual standard deviation is about 12.815, whereas the mean residual is about 1.1e-4.

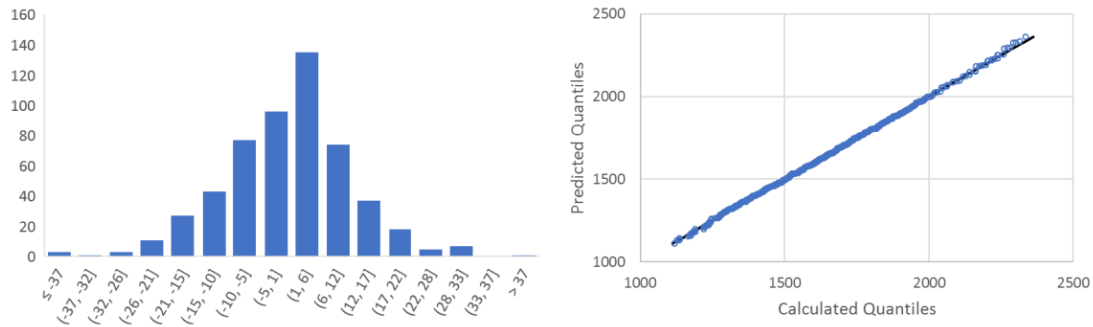

**Fig. 15. Goodness of the regression model of the wing-tip displacement response: residual distribution (left) and Q-Q plot (right)**

Concerning the model for the prediction of the wing structural mass, the analysis of DoE data again reveals the high influence of the usual variables, that is x1, x2, x4, x9, x13 and x15, whereas all other parameters have intermediate or low effect. For such model no evidence of quadratic behaviour of the response surface appears from the data; thus, pure quadratic terms are neglectable.

Let us define $m_{FW+RW}$ as the sum of the structural masses of the front and rear wings; as for the other models, the mass model has the following expression:

$$m_{FW+RW}(\boldsymbol{x}) \cong m_0 + \sum m_l \cdot x_l + \sum n_{lm} \cdot x_l \cdot x_m \qquad (11)$$

with model coefficients shown in Table 10. As already said, the regression model does not contain pure quadratic terms and includes only four 2-factor interaction effects. Therefore, the term $m_0$ represents the average main wings structural mass of the reference PrP-300 configuration. The mass estimated by means of Eq. (11) is expressed in kg.

**Table 10. Coefficients of the regression model for the wings mass prediction**

| Model coefficients | | | | | |
|---|---|---|---|---|---|
| $m_0$ | 16514.7 | $m_6$ | 173.0 | $n_{45}$ | 111.6 |
| $m_1$ | 686.6 | $m_8$ | 163.0 | $n_{46}$ | 112.0 |
| $m_2$ | 690.9 | $m_9$ | 573.5 | $n_{48}$ | 103.0 |
| $m_4$ | 756.2 | $m_{13}$ | 241.8 | $n_{915}$ | 39.0 |
| $m_5$ | 172.4 | $m_{15}$ | 222.0 | -- | -- |

Fig. 16 shows the Q-Q plot of the wings structural mass response; as we can see the predicted quantiles are very close to the theoretical ones, showing the goodness of the fitted model.

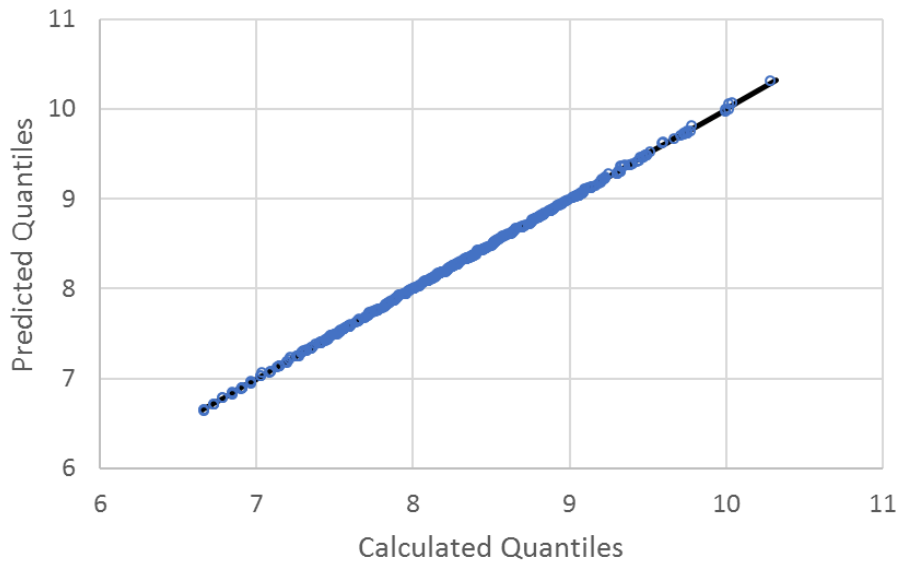

**Fig. 16. Q-Q plot of the wings structural mass response**

### 4.4    Optimization by means of DoE approach

The regression models obtained in the previous section are used to optimize the box-wing structure; the optimization problem is formulated as follows:

$$\begin{cases} min\big(m_{FW+RW}(\boldsymbol{x})\big) \\ \sigma_{VM}^{F}(\boldsymbol{x}) - \dfrac{\sigma_Y}{SF} \leq 0 \\ \sigma_{VM}^{R}(\boldsymbol{x}) - \dfrac{\sigma_Y}{SF} \leq 0 \\ u_z^{max(x)} - 0.1 \cdot \dfrac{b}{2} \leq 0 \\ -1 \leq x_i \leq 1 \end{cases} \qquad (12)$$

which provides the minimum wings structural mass under constraints on the stress in the main wings and the maximum wing-tip deflection. More precisely, the equivalent stress must not be greater than the admissible one, the latter given as the yield stress divided by an appropriate safety factor. If different materials are used, the lowest yield strength is considered. The constraint on the wing-tip deflection reflects the condition of geometric linearity, assumed to be valid if the wing-tip deflection is not greater of the 10% of the wing semi-span. Finally, the last condition of Eq. (12) defines the ranges of variation of the design variables. The optimization problem has been implemented in Python language ([41]) using the optimization module of Scipy ([42]). The optimal solution is searched by means of the Trust-Region Constrained algorithm.

The optimization problem given in Eq. (12) has been solved by adopting the DoE approach, considering the PrP-300 configuration, whose mass breakdown is summarized in Table 11. Therefore, the optimisation has been performed in order to minimize the mass of box-wing structures, while keeping the other masses unchanged.

**Table 11. Mass breakdown of the initial PrP-300 configuration**

|  | **Mass [kg]** | **Notes** |
|---|---|---|
| **Empty operative mass** | 72474 | - |
| **- Fuselage Structures** | 11230 | including openings and special structures according to [43] |
| **- Box-Wing Structures** | 17699 | including wing carry-trough structures and secondary structures |
| **- Vertical Tail Structures** | 1129 | - |
| **- Engines** | 13676 | Includes 2 bare engines with nacelles and pylons |
| **- Landing Gears** | 4500 | includes main and nose landing gear |
| **- Systems** | 10014 | includes hydraulic, pneumatic, anti-ice, electric, air condition and flight control systems; APU; Avionics; Instruments |
| **- Operatings** | 14180 | includes furnishing, oil, crew, crew seats, aircraft documents, pax seats, catering, emergency equipment, toilet fluids |
| **Fuel mass** | 27000 | harmonic point fuel |
| **Payload mass** | 29260 | payload equal to 308 passengers |
| **MTOW** | **128688** |  |

A safety factor equal to 1.2 has been applied to the yield strength of 345 MPa, thus the maximum equivalent stress in each wing has been limited to 287.5 MPa. In addition, according to Eq. (12), a maximum wing-tip deflection of 1800 mm has been prescribed. Finally, the upper and lower boundaries shown in Table 5 have been adopted.

Table 12 shows the main results of the structural optimization compared with those related to the initial PrP-300 configuration; focusing only on the box-wing structure, the mass saving is about 2100 kg, thus reducing the box-wing structural mass of about 12% and the maximum take-off mass of about 1.5%. Further details about these results are given in Section 5.1.

**Table 12. Results of the DoE-based optimization and comparison with the initial PrP-300 configuration**

|  | PrP-300 configuration | DoE-based optimization |
|---|---|---|
| **Box-Wing Structures** | 17699 | 15585 |
| **- Front Wing** | 9148 | 8094 |
| **- Rear Wing** | 7958 | 6898 |
| **- Vertical Wings** | 593 | 593 |
| **Empty operative mass [kg]** | 72474 | 70314 |
| **MTOW [kg]** | 128688 | 126574 |

## 5    Validation and comparison with existing models

### 5.1    *Optimum search validation*

The first aspect object of validation is the optimum search strategy implemented through the DoE. For such purpose, the result of the DoE-based optimization has been compared to another achievement of the PARSIFAL project, consisting in a wing structures optimization carried out from the same starting point, the PrP-300 configuration, using the optimisation code MIDACO ([44]), which implements a derivative-free evolutionary algorithm known as Ant Colony Optimization (ACO). Such solution has been obtained by maximizing the Von Mises stress in front wing structures, since in the typical solutions obtained in the aforementioned research projects concerning the PrP, the rear wing undergoes a lower wing loading in order to fulfil longitudinal stability and equilibrium requirements ([45]). Skins and stringers thickness have been indicated as optimization parameters, whereas all the other parameters defined in Table 3 have been set to their minimum values, as Table 13 summarizes. The choice of reducing the number of free parameters from 9 to 3 derives from the weak influence of the last 6 parameters in the list; this can be deduced from the analysis of DoE data provided in Section 4.3 regarding the stress acting on the front wing.

**Table 13. Setting of design parameters in MIDACO**

| Parameters | Type | Value(s) | |
|:---:|:---:|:---:|:---:|
| $t_{sk}^{FB}$ | variable | 6 | 14 |
| $t_{sk}^{FT}$ | variable | 6 | 14 |
| $t_{st}^{F}$ | variable | 1.5 | 7 |
| $h_{st}^{FT}$ | constant | 40 | |
| $h_{st}^{FB}$ | constant | 40 | |
| $t_{sk}^{RT}$ | constant | 6 | |
| $h_{st}^{RT}$ | constant | 40 | |
| $\alpha_{st}^{F}$ | constant | 0.7 | |
| $\tau_{sk}^{R}$ | constant | 0.25 | |

Starting from the given set of design parameters, the solution has been achieved by performing a FEM-based optimization aiming at minimizing the wing mass while fulfilling the constraint described by Eq. (13), where $\sigma_{eq,\,adm}$ is the admissible equivalent stress calculated according to von Mises criterion, $\sigma_{eq,\,max}$ is the yielding stress of the selected material and $SF$ is the prescribed safety factor.

$$\sigma_{eq,\,adm} \leq \frac{\sigma_{eq,\,max}}{SF} \qquad (13)$$

The solution obtained by means of the MIDACO solver, shown in Table 14, Fig. 17 and Fig. 18, has been then searched using the DoE-based approach, starting from the same initial point indicated in Table 13 and implementing the optimization problem shown in Eq. (12). Whereas in MIDACO only 3 design parameters have been used as optimization variables, in this case all of them have been considered as variables assigning the boundaries indicated in Table 13. Although the MIDACO optimization has been obtained with constraints acting only on the front wing, it has been observed that the extension of the same constraints to the rear wing does not affect the results. This behaviour is not general but typical of the box-wing, since it depends on the different wing loading of the two horizontal wings, which - as previously said and discussed in [45] – has to be lower for the rear wing. The combination of lower aerodynamic loads and minimum values assigned to structural parameters, which cannot be as small as desired because of manufacturing reasons, makes the rear wing not critical. For this reason, additional constraints acting on the rear wing do not affect the optimal solution. The results achieved with the DoE-based approach are shown in Table 14, where a comparison with MIDACO solution is given.

Looking at Table 14, it can be seen how the optimized solution is characterized by rear wing parameters assuming their lower boundary values, which is a result of the fact that stresses, although maximized, do not exceed the limits given by Eq.

(13). Since MIDACO optimization is carried out assigning to rear wing parameters the same constant values (see Table 13), the masses of the box-wing structure are very close to each other with a maximum difference < 1% for the front wing structures. Comparing the stress distribution achieved by the two wings within the MIDACO and DoE-based optimizations, Table 14, shows that the predicted maximum values are very close, whereas Fig. 17 and Fig. 18 indicate that similar stress fields are obtained.

**Table 14. Comparison between MIDACO and DoE-based optimizations**

| | PrP-300 configuration | MIDACO optimization | DoE-based optimization | Δ% (1-DoE/MIDACO) |
|---|---|---|---|---|
| $\sigma_{eq,\,max}$[MPa] | 345 | 345 | 345 | - |
| SF [ - ] | 1.2 | 1.2 | 1.2 | - |
| $t_{sk}^{FB}$[mm] | 14 | 9.7 | 10.3 | 6.2% |
| $t_{sk}^{FT}$[mm] | 10 | 8.0 | 8.3 | 3.8% |
| $t_{st}^{F}$[mm] | 8 | 6.9 | 5.7 | -17.4% |
| $h_{st}^{RT}$[mm] | 50 | 40.0 | 40.0 | 0.0% |
| $h_{st}^{FB}$[mm] | 50 | 40.0 | 40.0 | 0.0% |
| $t_{sk}^{RT}$[mm] | 8 | 6.0 | 6.0 | 0.0% |
| $h_{st}^{RT}$[mm] | 50 | 40.0 | 40.0 | 0.0% |
| $\alpha_{st}^{F}$[ - ] | 1 | 0.7 | 0.7 | 0.0% |
| $\tau_{sk}^{R}$[ - ] | 0.49 | 0.25 | 0.25 | 0.0% |
| Front wing max. stress (von Mises) [MPa] | 229 | 287.5 | 285 | -0.9% |
| Rear wing max. stress (von Mises) [MPa] | 188 | 259 | 254 | -1.9% |
| Front Wing tip deflection [mm] | 1470 | 1804 | 1794 | -0.6% |
| Front Wing structural mass [kg] * | 9148 | 8154 | 8094 | -0.7% |
| Rear Wing structural mass [kg] * | 7958 | 6898 | 6898 | 0.0% |
| Vertical Wing structural mass [kg] * | 593 | 593 | 593 | 0.0% |
| Box-Wing structural mass [kg] * | 17699 | 15645 | 15585 | 0.0% |
| * Including secondary structures | | | | |

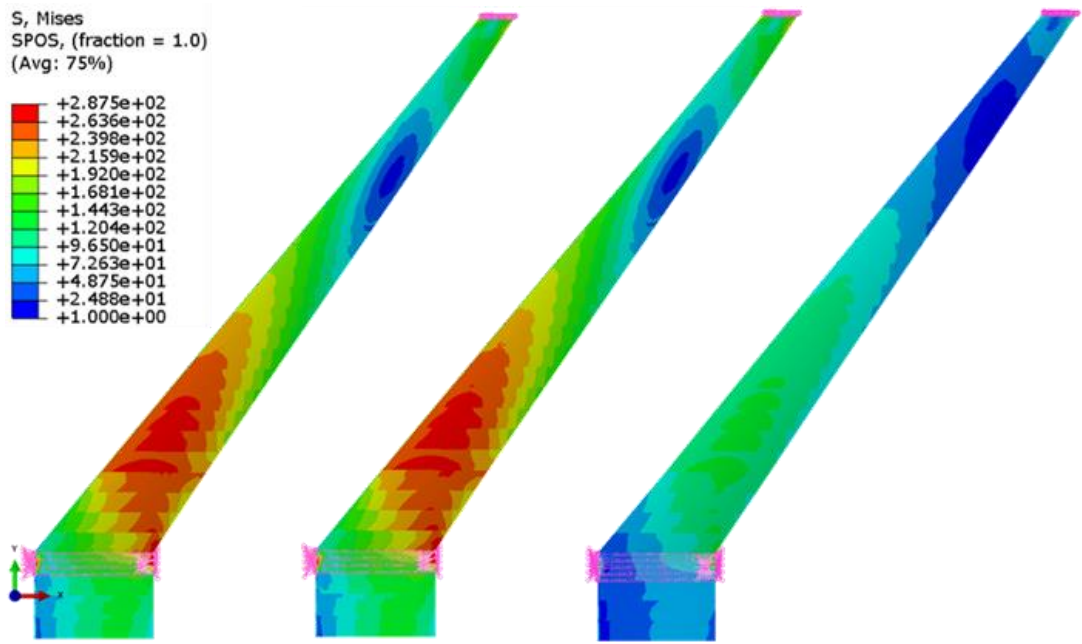

**Fig. 17. Von Mises stress in the front wing (upper skin): DoE-based optimization (left), MIDACO optimization (centre) and PrP-300 configuration (right)**

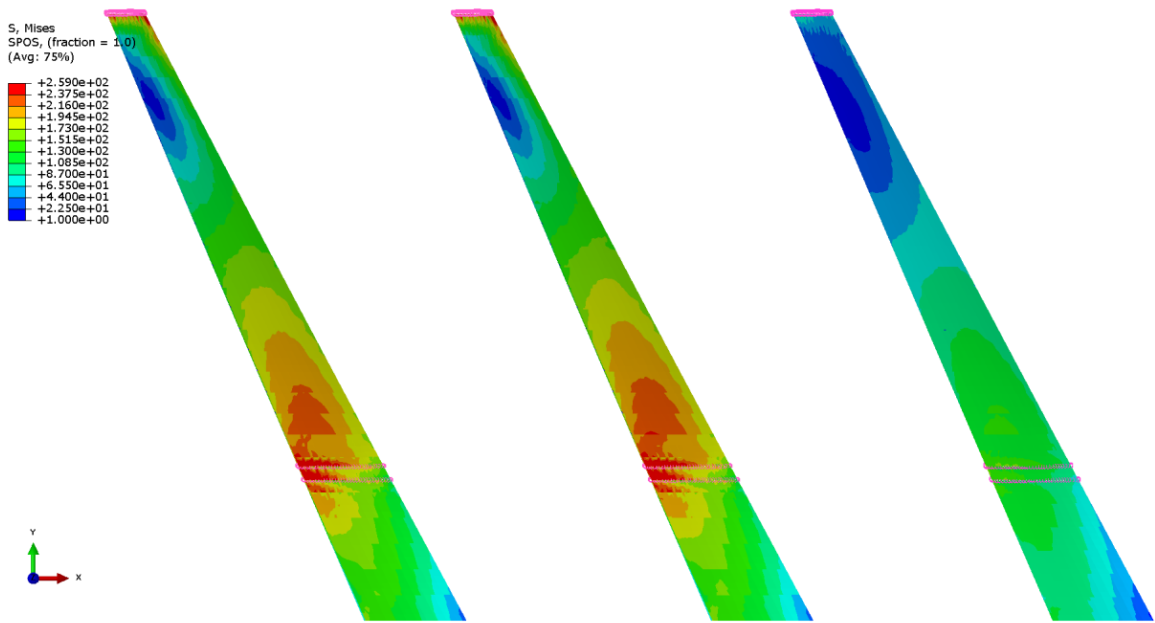

**Fig. 18. Von Mises stress in the rear wing (upper skin): DoE-based optimization (left), MIDACO optimization (centre) and PrP-300 configuration (right)**

It is worth to remark that the solution obtained takes the different roles that front and rear wing play into account, showing that the maximum stress constraint ($\sigma_{VM} \leq 285$ MPa) acts only on the front wing. Since the wing loading is a result of the VLM simulation, hence depending on box-wing geometry and flight condition, it is possible to conclude that the wing mass estimation introduced by this approach is physic based, hence influenced by the box-wing aerodynamic and flight mechanic characteristics, including lifting surfaces geometry.

*5.2    Comparisons with existing models*

Once the optimum search strategy has been validated, the proposed approach has been adopted to a case available from the literature ([15]), according to which the mass of each wing composing the box-wing system can be evaluated by means of the formula given in Eq. (14), where b is the wingspan, S is the gross wing area, $\Lambda_{1/4}$ is the quarter chord sweep angle, $\lambda$ is the taper ratio, $n_z$ is the vertical load factor, $M_{TOM}$ is the maximum take-off mass, $V_D$ is the diving speed and t/c is the average thickness-to-chord ratio.

$$M_W = 0.028 \left[ \frac{bS}{cos\,\Lambda_{1/4}} \left( \frac{1+2\lambda}{1+3\lambda} \right) \left( \frac{n_z M_{TOM}}{S} \right)^{0.3} \left( \frac{V_D}{t/c} \right)^{0.5} \right]^{0.9} \tag{14}$$

Such formula can be applied to the front wing of the "PrP-300" configuration, using the values provided in Table 15, which are referred to the equivalent simply tapered wing derived according to [46] from the actual design which shows the presence of a kink.

The result of the application of Eq. (14) to the front wing is presented in Table 15 where DoE-based optimization results are also shown. In order to use the same assumptions of cases which Eq. (14) derives from, the safety factor (SF) applied is 1.5. As shown, the difference in terms of mass estimation between the two cases is below 2%, which provides a further validation of the method here proposed.

In the case of the rear wing, the comparison between the estimated mass provided by the DoE-based optimization and Eq. (14) is shown in Table 16. Unlike the previous case where the Jemitola's formula provides a lower mass, the present model predicts a rear wing mass 272 kg lower than the Jemitola's formula. The difference between the two cases however is below 4%. It is interesting to note that if we consider the sum of the masses of the two wings the difference is reduced to 160 kg corresponding to 1% of the total mass of the two wings.

**Table 15. Input and output of the comparison between the DoE-based optimization and Jemitola's formula for the front wing mass prediction**

| Wing data (front wing) | | DoE-based optimization (kinked wing) | Jemitola's formula (equivalent wing) |
|---|---|---|---|
| **INPUT** | **S [m²]** | 161.48 | 153.14 |
| | **b [m]** | 35 | 35 |
| | $\Lambda_{1/4}$ **[deg]** | Inboard: 33 / Outboard: 38 | 38.1 |
| | $\lambda$ **[-]** | Inboard: 0.74 / Outboard: 0.36 | 0.29 |
| | **t/c [-]** | 0.11 | 0.11 |
| | $\sigma_{eq,\,max}$**[MPa]** | 345 | N/A |
| | **SF [ - ]** | 1.5 | N/A |
| | $M_{TOM}$**[kg]** | 126414 | 126414 |
| | $n_z$**[-]** | 2.5 | 2.5 |
| | $V_D$ **[m/s]** | N/A | 245 |
| **OUTPUT** | $t_{sk}^{FB}$**[mm]** | 12.7 | N/A |
| | $t_{sk}^{FT}$**[mm]** | 12.6 | N/A |
| | $t_{st}^{F}$**[mm]** | 3.9 | N/A |
| | $h_{st}^{FT}$**[mm]** | 40.0 | N/A |
| | $h_{st}^{FB}$**[mm]** | 40.0 | N/A |
| | $t_{sk}^{RT}$**[mm]** | 6 | N/A |
| | $h_{st}^{RT}$**[mm]** | 40 | N/A |
| | $\alpha_{st}^{F}$**[ - ]** | 0.70 | N/A |
| | $\tau_{sk}^{R}$**[ - ]** | 0.25 | N/A |
| | **Wing mass [kg]** | 8982 | 8870 |

**Table 16. Input and output of the comparison between the DoE-based optimization and Jemitola's formula for the rear wing mass prediction**

| Wing data (rear wing) | | DoE-based optimization | Jemitola's formula |
|---|---|---|---|
| **INPUT** | $S$ [m²] | 127.5 | 127.5 |
| | $b$ [m] | 35 | 35 |
| | $\Lambda_{1/4}$ [deg] | -24.2 | -24.2 |
| | $\lambda$ [-] | 0.38 | 0.38 |
| | $t/c$ [-] | 0.11 | 0.11 |
| | $\sigma_{eq,\,max}$[MPa] | 345 | N/A |
| | SF [ - ] | 1.5 | N/A |
| | $M_{TOM}$[kg] | 126414 | 126414 |
| | $n_z$[-] | 2.5 | 2.5 |
| | $V_D$ [m/s] | N/A | 245 |
| **OUTPUT** | $t_{sk}^{FB}$[mm] | 12.7 | N/A |
| | $t_{sk}^{FT}$[mm] | 12.6 | N/A |
| | $t_{st}^{F}$[mm] | 3.9 | N/A |
| | $h_{st}^{FT}$[mm] | 40.0 | N/A |
| | $h_{st}^{FB}$[mm] | 40.0 | N/A |
| | $t_{sk}^{RT}$[mm] | 6 | N/A |
| | $h_{st}^{RT}$[mm] | 40 | N/A |
| | $\alpha_{st}^{F}$[ - ] | 0.70 | N/A |
| | $\tau_{sk}^{R}$[ - ] | 0.25 | N/A |
| | **Wing mass [kg]** | 6900 | 7172 |

Therefore, it can be concluded that the proposed model is in good agreement with the Jemitola's formula for estimating the wing structural masses of a box-wing structure and that the maximum difference between the two methods is below 4%.

# 6    Application to regional box-wing aircraft design

As said before, the development of a DoE-based optimisation approach has been made possible also thanks to the results achieved within the H2020 research project PARSIFAL, concerning the application of the box-wing architecture to short-to-medium haul aircraft. In order to demonstrate the applicability of such approach outside the framework it has been conceived, a box-wing regional aircraft object of study in the Italian research project "PROSIB" has been taken into account.

The project PROSIB, carried out between 2018 and 2021, concerns the study of hybrid-electric propulsion systems applied to commuter aircraft ([47]) as well as to regional aircraft with both conventional ([48]) and disruptive architectures. Concerning these latter, the project focuses on box-wing aircraft design as well as on the development of design tools and methodologies for such purpose ([29]). As Table 17 shows, a box-wing 40 pax regional aircraft has been designed, also considering the possible integration of a Distributed Electric Propulsion system on the front wing.

**Table 17. TLARs and dimensions of the regional box-wing aircraft studied within the project PROSIB**

| | | |
|---|---|---|
| **Max n° passengers** | 40 | 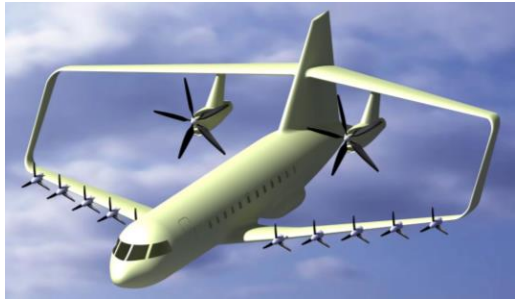 |
| **Seat abreast** | 2-2 | |
| **Design Range** | 600 nm | |
| **Cruise altitude** | FL 200 | |
| **Cruise speed** | 275 kts | |
| **Ref. Wing Area** | 64.9 m$^2$ | |
| **Wingspan** | 22.1 m | |
| **Fuselage length** | 21.9 m | |
| **MTOW** | 21148 kg | |

Although this configuration is not the arrival point of the project, it has been object of several analyses, such as:

- VLM-driven optimization through aforementioned code "AEROSTATE";

- mission analysis through the in-house developed code "THEA-CODE" ([29]);

- aerodynamic model validation through CFD analyses;

- structural weight estimation from empiric models (such as in [37]) and following refinement through FEM analyses;

- inertia moments estimation from FEM analyses;

- propellers design from CFD analyses carried out by the project partner CIRA (Centro Italiano Ricerche Aerospaziali);

- sizing of control surfaces and take-off dynamic simulation.

Therefore, given the large amount of available data, such configuration has been adopted a further test case for the development of the DoE based approach proposed in this work.

*6.1    DoE implementation and results*

The first main step consists in defining the design space, i.e. the set of the relevant structural parameters which has to be used to build the DoE dataset. Table 18 provides the complete design parameters and their boundaries adopted in this case. The experience gained for the PARSIFAL case has provided the guidelines to achieve a simplified parametrization of the structural model; at the same time new design parameters for the rear wing have been introduced, leading, thus, to a design space more representative of the whole box-wing architecture. In particular, the design variables related to the spanwise distribution of the skin thickness and stringer's dimensions have been disregarded, since a weak relationship between these parameters and the regression models has been observed in the PARISFAL case. Similar considerations also apply to the stringer's height which has now been assumed to be the same for both wings and both skins of each wing, whereas the thickness of both skins of the rear wing, as well as the thickness of the stringers and the spar's webs are now considered.

Lower and upper boundaries of each design parameters have been set according to previous results achieved within the project PROSIB.

**Table 18. Setting of design variables and ranges of variation**

|  | Parameters | Lower boundaries | Upper boundaries | Step size | Codified variables |
|---|---|---|---|---|---|
| Front Wing | $t_{sk}^{FT}$ [mm] | 1.25 | 5.00 | 1.875 | x1 |
|  | $t_{sk}^{FB}$ [mm] | 1.25 | 5.00 | 1.875 | x2 |
|  | $t_{web}^{F}$ [mm] | 1.50 | 6.50 | 2.5 | x3 |
|  | $t_{st}^{F}$ [mm] | 1.50 | 6.50 | 2.5 | x4 |
| Rear Wing | $t_{sk}^{RT}$ [mm] | 1.25 | 5.00 | 1.875 | x5 |
|  | $t_{sk}^{RB}$ [mm] | 1.25 | 5.00 | 1.875 | x6 |
|  | $t_{web}^{R}$ [mm] | 1.50 | 6.50 | 2.5 | x7 |
|  | $t_{st}^{R}$ [mm] | 1.50 | 6.50 | 2.5 | x8 |
| Any wing / Any panel | $h_{st}$ [mm] | 30.00 | 45.00 | 7.5 | x9 |

In this case, the same approach described in Section 4.3 has been applied in order to build the regression models for the prediction of the mass of the box-wing structure, the stress acting in each wing and the wing-tip deflection. Such models maintain the same expression given by Eq. (8) – Eq. (11), whereas the coefficients of each model are shown in Table 19 – Table 22.

**Table 19. Coefficients of the regression model for the stress response in the front wing**

| Model coefficients | | | | | | | |
|---|---|---|---|---|---|---|---|
| $a_0^F$ | 177.383 | $b_8^F$ | -8.697 | $c_{23}^F$ | 3.943 | $c_{44}^F$ | 7.169 | $d_{124}^F$ | 6.996 |
| $b_1^F$ | -48.121 | $b_9^F$ | -15.516 | $c_{24}^F$ | 9.958 | $c_{45}^F$ | 2.085 | $d_{125}^F$ | 1.167 |
| $b_2^F$ | -36.301 | $c_{11}^F$ | 39.5 | $c_{25}^F$ | 1.498 | $c_{46}^F$ | 2.039 | $d_{126}^F$ | 1.253 |
| $b_3^F$ | -21.762 | $c_{12}^F$ | -28.631 | $c_{26}^F$ | 1.556 | $c_{48}^F$ | 2.719 | $d_{128}^F$ | 1.465 |
| $b_4^F$ | -46.351 | $c_{13}^F$ | 6.107 | $c_{28}^F$ | 1.887 | $c_{49}^F$ | -1.268 | $d_{129}^F$ | 2.19 |
| $b_5^F$ | -6.931 | $c_{14}^F$ | 13.243 | $c_{29}^F$ | 2.873 | $c_{58}^F$ | 1.789 | $d_{134}^F$ | -2.47 |
| $b_6^F$ | -6.93 | $c_{19}^F$ | 4.108 | $c_{34}^F$ | 9.34 | $c_{68}^F$ | 1.736 | $d_{234}^F$ | -2.333 |
| $b_7^F$ | -2.194 | $c_{22}^F$ | 30.156 | $c_{39}^F$ | 2.404 | $d_{123}^F$ | 6.502 | $e_{1234}^F$ | -3.4 |

**Table 20. Coefficients of the regression model for the stress response in the rear wing**

| Model coefficients | | | | | | | |
|---|---|---|---|---|---|---|---|
| $a_0^R$ | 161.686 | $b_8^R$ | - | $c_{25}^R$ | 1.626 | $c_{56}^R$ | -23.139 | $c_{78}^R$ | 7.93 | $d_{456}^R$ | 1.51 |
| $b_1^R$ | -10.766 | $b_9^R$ | - | $c_{26}^R$ | 1.872 | $c_{57}^R$ | 5.879 | $c_{79}^R$ | 2.2 | $d_{458}^R$ | -1.462 |
| $b_2^R$ | -11.462 | $c_{14}^R$ | 2.526 | $c_{28}^R$ | 2.031 | $c_{58}^R$ | 10.982 | $c_{88}^R$ | 11.358 | $d_{567}^R$ | 4.764 |
| $b_3^R$ | -3.047 | $c_{15}^R$ | 1.575 | $c_{29}^R$ | 1.627 | $c_{59}^R$ | 3.697 | $c_{89}^R$ | -1.327 | $d_{568}^R$ | 9.635 |
| $b_4^R$ | -11.23 | $c_{16}^R$ | 1.64 | $c_{45}^R$ | 1.689 | $c_{66}^R$ | 22.634 | $d_{156}^R$ | 1.471 | $d_{569}^R$ | 3.1 |
| $b_5^R$ | -38.5 | $c_{18}^R$ | 1.903 | $c_{46}^R$ | 1.577 | $c_{67}^R$ | 2.997 | $d_{158}^R$ | -1.246 | $d_{578}^R$ | -2.5 |
| $b_6^R$ | -30.88 | $c_{19}^R$ | 1.507 | $c_{48}^R$ | 1.837 | $c_{68}^R$ | 12.586 | $d_{256}^R$ | 1.689 | $d_{678}^R$ | -1.777 |
| $b_7^R$ | -18.149 | $c_{24}^R$ | 2.657 | $c_{55}^R$ | 28.183 | $c_{69}^R$ | 3.703 | $d_{258}^R$ | -1.378 | $e_{5678}^R$ | -2.609 |

**Table 21. Coefficients of the regression model for the wing-tip deflection response**

| Model coefficients | | | | | | | | | |
|---|---|---|---|---|---|---|---|---|---|
| $A_0$ | 500.339 | $C_{11}$ | 18.924 | $C_{24}$ | 19.196 | $C_{45}$ | 3.936 | $C_{67}$ | 4.424 |
| $B_1$ | -61.178 | $C_{13}$ | 5.738 | $C_{25}$ | 4.863 | $C_{46}$ | 4.039 | $C_{68}$ | 16.282 |
| $B_2$ | -66.144 | $C_{14}$ | 18.262 | $C_{26}$ | 4.831 | $C_{47}$ | 0.812 | $C_{69}$ | 5.593 |
| $B_3$ | -22.628 | $C_{15}$ | 3.902 | $C_{27}$ | 1.172 | $C_{48}$ | 5.318 | $C_{78}$ | 8.342 |
| $B_4$ | -67.484 | $C_{16}$ | 3.894 | $C_{28}$ | 6.071 | $C_{49}$ | -4.261 | $C_{79}$ | 2.437 |
| $B_5$ | -50.568 | $C_{17}$ | 0.981 | $C_{29}$ | 6.628 | $C_{55}$ | 16.695 | $C_{88}$ | 15.355 |
| $B_6$ | -48.214 | $C_{18}$ | 4.962 | $C_{34}$ | 8.672 | $C_{57}$ | 6.602 | $C_{89}$ | -3 |
| $B_7$ | -21.338 | $C_{19}$ | 5.99 | $C_{38}$ | 0.586 | $C_{58}$ | 17.159 | $D_{134}$ | -3.564 |
| $B_8$ | -63.95 | $C_{22}$ | 19.257 | $C_{39}$ | 2.257 | $C_{59}$ | 5.837 | $D_{234}$ | -3.702 |
| $B_9$ | -37.066 | $C_{23}$ | 6.504 | $C_{44}$ | 12.654 | $C_{66}$ | 15.628 | $D_{578}$ | -3.465 |

**Table 22. Coefficients of the regression model for the wings mass prediction**

| Model coefficients | | | | | |
|---|---|---|---|---|---|
| $m_0$ | 2158 | $m_4$ | 110 | $m_8$ | 105.4 |
| $m_1$ | 89.8 | $m_5$ | 67.4 | $m_9$ | 77.2 |
| $m_2$ | 89.8 | $m_6$ | 67 | $n_{49}$ | 24.6 |
| $m_3$ | 69.2 | $m_7$ | 43.4 | $n_{89}$ | 23.6 |

The optimization problem is still solved by means of Eq. (12); Table 23 provides a summary in term of the design parameters, assumed initial values and the resulting values, obtained through the DoE-based optimization approach.

**Table 23. Results of the DoE-based optimization**

| Parameters | Type | Initialization | Optimization output |
|---|---|---|---|
| $t_{sk}^{FB}$ [mm] | variable | 3.12 | 4.79 |
| $t_{sk}^{FT}$ [mm] | variable | 3.12 | 3.47 |
| $t_{web}^{F}$ [mm] | variable | 4.00 | 1.50 |
| $t_{st}^{F}$ [mm] | variable | 4.00 | 1.93 |
| $t_{sk}^{RB}$ [mm] | variable | 3.12 | 2.63 |
| $t_{sk}^{RT}$ [mm] | variable | 3.12 | 3.81 |
| $t_{web}^{R}$ [mm] | variable | 4.00 | 1.50 |
| $t_{st}^{R}$ [mm] | variable | 4.00 | 2.80 |
| $h_{st}^{FT}$ [mm] | variable | 37.5 | 30.00 |
| $h_{st}^{FB}$ [mm] | variable | 37.5 | 30.00 |
| $h_{st}^{RT}$ [mm] | variable | 37.5 | 30.00 |
| $h_{st}^{RB}$ [mm] | variable | 37.5 | 30.00 |

In this case, the achieved solution has been compared to the results achieved within the PROSIB project, taking both the preliminary estimations, performed according to the well-known empirical models described in [37], and higher fidelity results, obtained in latter stages of the project by means of FEM analyses. As Table 24 shows, DoE-based optimization allows to achieve, at a lower computational cost, results that are very close to FEM predictions, with errors on predicted stresses below 10%. Being developed for conventional cantilever wings, empirical models show their limitations providing an overestimation of the wing mass.

**Table 24. Comparison of results obtained from the DoE-based optimization, FEM analyses and empirical model**

|  | DoE-based optimization | FEM analyses | Empirical models ([37]) |
|---|---|---|---|
| **Input: $\sigma_{eq,\,max}$[MPa]** | 345 | 345 | N/A |
| **Input: SF [ - ]** | 1.5 | 1.5 | N/A |
| **Box-Wing structural mass [kg]** | 2109.2 | 2112.2 | 3018.5 |
| **- Front wing [kg]** | 1101.8 | 1117.2 | 1593.7 |
| **- Rear wing [kg]** | 859.8 | 847.4 | 1256.7 |
| **- Vertical Wings [kg]** | 147.6 | 147.6 | 168.07 |
| **Front wing max. stress (von Mises) [MPa]** | 230 | 214.3 | - |
| **Rear wing max. stress (von Mises) [MPa]** | 230 | 216.4 | - |
| **Front Wing tip deflection [mm]** | 600 | 580.2 | - |

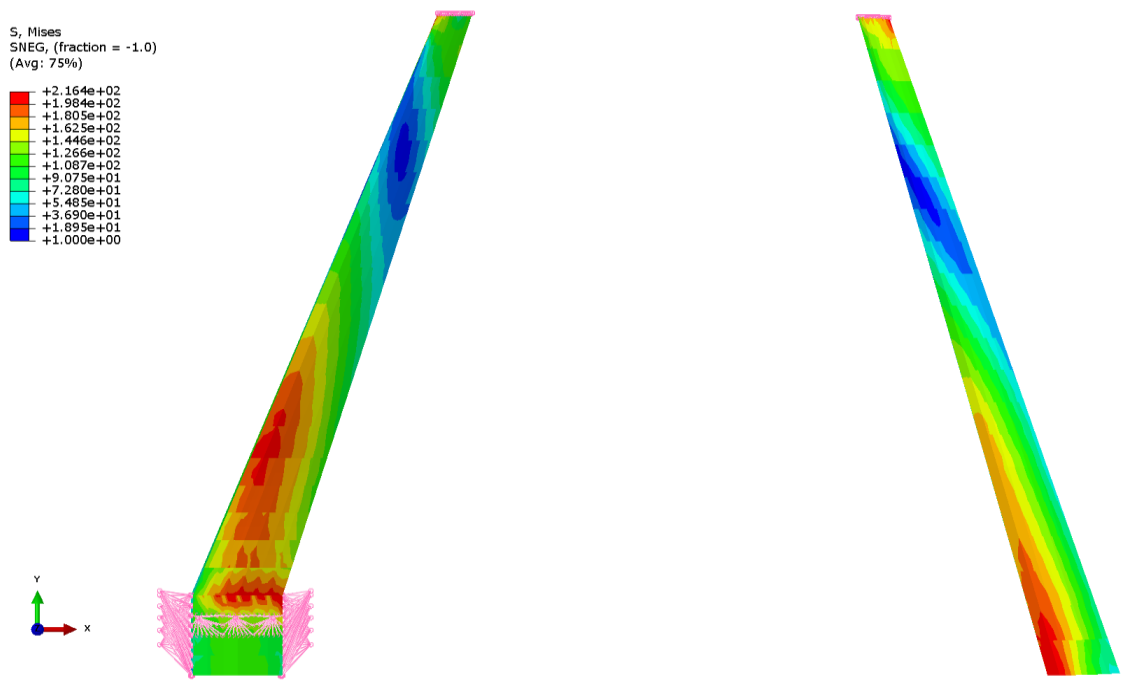

**Fig. 19. Von Mises stress in the regional aircraft configuration**

In addition, the comparison between the stress fields shown in Fig. 19 show the capability of this approach to define local behaviour of the structures. For such reasons, if used in the preliminary design phases, such approach allows to overcome the limits of the empirical models, without big penalties in terms of computational cost.

# 7    Conclusions and further development

Aiming to introduce a structural mass prediction model suitable for the early stages of the design of box-wing aircraft, the present paper presents the implementation and the outcomes of the application of a Design of Experiment (DoE) approach for the definition of regression models useful for the declared scope.

The paper illustrates how regression models for the Von Mises stresses, wing-tip displacements and total wings' structural mass can be built starting from a dataset of FEM results, generated after the definition of a parametric structural model of the aircraft object of interest.

The case of the box-wing aircraft "PrP-300" resulting from the project PARSIFAL is taken as the main reference to show this and the following steps of the method. The set of parameters describing the "PrP-300" structural model is presented and the possible assumptions useful to reduce the number of these parameters are introduced. As described, this operation is fundamental to reduce the complexity of the problem as well as the computational cost of the approach, since the FEM database construction needs more than $2^n$ evaluations, with $n$ indicating the number of parameters. As a result, the number of parameters is reduced from 17 to 9.

The FEM analyses are then described, underlining the role of the parametric pre-processor "WAGNER", developed within the PARSIFAL project in order to manage both conventional and box-wing architectures, and the details concerning the FE model adopted for structural analyses with the code ABAQUS. As sizing condition, the static load case associated to +2.5 load factor has been considered, taking also pressure loads and engines' thrust into account. Given the admissible stress calculated considering aluminium alloys Al-2024 and Al-7075 as materials, a sizing criterion based on the limitations of the maximum equivalent stress has been implemented in the FEM database construction procedure.

With the FEM database available, the DoE approach is introduced, focusing on the sensitivity analysis strategy that allows to recognize the most important structural parameters, i.e. to quantify the coefficients to be associated to each design parameter, or a combination of them, within the regression models. Details on the sensitivity analyses are provided for each regression model, and results are given in terms of sets of coefficients for the prediction of stress response in front and rear wings, wing-tip displacement and total wings' structural mass.

Given the PrP-300 as the starting point, the regression models are used to solve an optimization problem in which the total wings' structural mass plays the role of the objective function, whereas stress response and wing-tip displacements are constrained quantities. As a result, the DoE-based optimization provides a reduction of structural mass of about 12%. In order to validate this result, two comparisons are carried out: the first one, aiming at validating the optimum search strategy, in which the mean of comparison is the result of an optimization carried out with the optimization code MIDACO, and the second one, in which the front wing sized according to the DoE-based optimization is compared to results given by an empirical approach which adapts mass estimation formulas for cantilever wings to the box-wing architecture ([15]). Looking at wings mass prediction, both the comparisons have shown errors below 2%.

Aiming to assess the DoE-based approach on a test case different from the one in which it has been developed, an additional test case has been considered. This is based on the preliminary results of the project PROSIB, which concerns the design of regional hybrid-electric aircraft with conventional or box-wing architecture. The same steps presented for the PrP-300 case are briefly outlined, showing the results and in particular the structural mass prediction from the DoE-based approach compared to direct FEM analyses and empirical models available in the literature for cantilever wings.

As for the PrP-300, the DoE-based optimization provides estimations as accurate as those from direct FEM analyses and allows to overcome the limitations of empirical models not conceived for box-wing architectures. The achieved results encourage the implementation of the regression models in the design loops that will be performed in the future activities related to both PARSIFAL and PROSIB projects. Such activity would assign a major role to the structural sizing also in the early stage design, thus bringing to a more comprehensive and multidisciplinary design environment.

As anticipated in Section 3, the method here presented allows to size wing structures adopting static strength as design criteria. As shown in [31], this simplification leads to acceptable underestimations (about 10%) of primary wing structures weight only for large aircraft with wing loading above 600 $kg/m^2$, whose design is strength dominated. For smaller aircraft the errors introduced when buckling effects are not taken into account are much higher. Therefore, it is worth underlining that the wing mass estimations reported in the paper are affected by some errors, which are certainly acceptable for the PARSIFAL case, whose wing loading is close to 600 $kg/m^2$, whereas would need further assessment for the PROSIB case.

In conclusion, the paper shows that the DoE approach provides a useful strategy to make the structural sizing more relevant in the preliminary design phases of box-wing aircraft, but every obtained solution would need to be verified and revised if necessary, according to the all the relevant design criteria. The parametric model here adopted for wing-box structures has been conceived to allow in the future a more comprehensive assessment of the box-wing structural behaviour, including important phenomena such as buckling. Therefore, additional cross-cutting development paths for the approach here presented concern the implementation of additional load cases and sizing criteria (e.g. buckling, fatigue, aeroelastic phenomena) in the FEM database construction, as well as the extension of the approach to structures in composite materials.

# 8    Acknowledgments

This work was supported by the European Union Horizon2020 Programme within the research project PARSIFAL (grant agreement n.723149) and by the Italian Ministry of Education, University and Research within the research project PROSIB (grant number ARS01_00297).